\documentclass{aa}
\usepackage{graphicx}
\usepackage{txfonts}
\usepackage{psfig}
\usepackage{psfrag}
\usepackage{latexsym}
\usepackage{times,amssymb,lscape,verbatim}
\usepackage{natbib}
\usepackage[mathcal]{euscript}
\usepackage{mathrsfs}
\begin{document}
   \title{Investigating grain growth in disks around southern T Tauri stars at millimetre wavelengths}


   \author{Dave Lommen\inst{1}
   	\and Chris M. Wright\inst{2}
	\and Sarah T. Maddison\inst{4}
	\and Jes K. J{\o}rgensen\inst{3}
	\and Tyler L. Bourke\inst{3}
	\and Ewine F. van Dishoeck\inst{1}
	\and Annie Hughes\inst{4}
	\and David J. Wilner\inst{3}
	\and Michael Burton\inst{5}
	\and Huib Jan van Langevelde\inst{6,1}
	     }

   \offprints{Dave Lommen, \\ \email{dave@strw.leidenuniv.nl}}

   \institute{Leiden Observatory, P.O. Box 9513, 2300 RA Leiden, The Netherlands
	\and School of Physical, Environmental and Mathematical Sciences, UNSW@ADFA, Canberra ACT 2600, Australia
	\and Harvard-Smithsonian Center for Astrophysics, 60 Garden Street, Cambridge, MA 02138, USA
	\and Centre for Astrophysics and Supercomputing, Swinburne University of Technology, PO Box 218, Hawthorn, VIC 3122, Australia
	\and School of Physics, University of New South Wales, Sydney, NSW 2052, Australia
	\and Joint Institute for VLBI in Europe, P.O. Box 2, 7990 AA Dwingeloo, The Netherlands
             }

   \date{Received ?? ; Accepted ??}

 \abstract 
  {Low-mass stars form with disks in which the coagulation of grains may eventually lead to the formation of planets. It is not known when 
    and where grain growth occurs, as models that explain
    the observations are often degenerate. A way to break this degeneracy is to resolve the sources under study.}
 {To find evidence for the existence of grains of millimetre sizes in disks around in T Tauri stars, implying grain growth.}
  {The Australia Telescope Compact Array (ATCA) was used to observe 15 southern
   T Tauri stars, five in the constellation Lupus and ten in 
   Chamaeleon, at 3.3~millimetre. The five Lupus sources were also observed with the
   Submillimeter Array (SMA) at 1.4~millimetre. Our new data are complemented with data from 
   the literature to determine the slopes of the spectral energy 
   distributions in the millimetre regime.}
{Ten sources were detected at better than 3$\sigma$ with the ATCA, with $\sigma \approx$ 1--2~mJy, and all
  sources that were observed with the SMA were detected at better than 15$\sigma$, with $\sigma \approx$ 4~mJy. 
  Six of the sources in our sample are resolved to physical radii of $\sim 100$~AU. Assuming that the emission from such large disks 
  is predominantly optically thin,
  the millimetre slope can be related directly to the opacity 
  index. For the other sources, the opacity indices are lower limits. Four out of six resolved sources have opacity indices $\lesssim 1$, indicating grain growth to 
  millimetre sizes and larger. The masses of the disks range from $< 0.01$ to 0.08
  M$_\odot$, which is comparable to the minimum mass solar nebula. A tentative correlation is found between the millimetre slope and the strength and shape 
  of the 10-$\mu$m silicate feature, indicating that grain growth occurs on similar (short) timescales in both the inner and outer disk.}
{}


   \maketitle
%

\section{Introduction}\label{sect: introduction}

   Disks of dust and gas are observed around many young stars. According to the so-called core-accretion model \citep{safronov:1969}, planetary systems such as our 
   own Solar System are formed in these circumstellar disks: the solid
   particles coagulate to form larger grains, which will grow to eventually form planets. The grains mainly consist of carbon and silicates.
   The silicate grains are readily observed through their 10- and 20-$\mu$m features. Both the change from amorphous to more crystalline grains, as well as the growth of grains 
   from submicron sizes to sizes of several $\mu$m have 
   been observed by the Infrared Space Observatory 
   \citep[ISO, ][]{malfait:1998, vanboekel:2005}, and more recently by the Spitzer Space Telescope \citep{kessler-silacci:2006}.
   Although the qualitative picture of grain growth has become much clearer over the last few years, several quantitative details are still under discussion. Open questions
   include for example the timescale over which grain growth occurs and how this relates to the disk's physical structure (e.g., its temperature and density profile). See
   \citet{dominik:2006} for a recent discussion of both laboratory measurements and theoretical modelling of the aggregation of dust in protoplanetary disks.
   
   A large sample of solar-mass T Tauri stars 
   have recently been observed with the InfraRed Spectrograph (IRS) on board the Spitzer Space 
   Telescope, in the context of the ``Cores to Disks'' (c2d) legacy \citep{evans:2003} and other programmes. 
   Most of the sources in the c2d sample show 10- and 20-$\mu$m amorphous silicate features \citep[][]{kessler-silacci:2006}, confirming the results of earlier ISO and 
   ground-based 10-$\mu$m observations \citep[e.g.,][]{vanboekel:2003, przygodda:2003}, and extending the observed protoplanetary disk sample to lower mass objects.
   The data indicate a large variety of silicate profiles, ranging 
   from strongly peaked silicate bands and steeply rising spectral energy distributions (SEDs) to ``boxy'' silicate profiles and flat SEDs. The boxy features with low 
   feature-to-continuum ratios are interpreted as grain growth to micron size \citep{bouwman:2001}.
   
   One possible explanation for the different spectra and SEDs is that grain growth and the shape of the disk are related. 
   \citet{dullemond:2004} identified a correspondence between the growth of grains in a circumstellar disk and the evolution of the disk from a ``flaring'' to a ``self-shadowed''
   geometry.
   In their models, the larger, more massive dust grains settle to the midplane of the disks as the grains 
   grow, and the initially flared disks evolve into flatter, self-shadowed disks. To what extent this process is related to the age of the young star is 
   still under debate; there are indications that some young stellar objects evolve more quickly than others. 
   
   Numerical models show that the process of settling and coagulation is rapid, being well underway at distances of 1--30~AU from the central star in $\sim 10^4$~yrs
   \citep[e.g.,][]{nomura:2006}. Models predict that the slopes of the SEDs in the millimetre wavelength range will become shallower as the grains in the disk grow to 
   millimetre (mm) and subsequently centimetre (cm) and larger sizes \citep{dullemond:2004, dalessio:2006}.
   It is therefore necessary to observe these sources at larger wavelengths than the infrared. Furthermore, millimetre observations probe the entire disk, including the 
   cold midplane, whereas infrared observations can only probe the hot surface layer of the inner disk.
   However, a shallow millimetre slope in itself is not enough evidence for grain growth in the disks, since an excess flux at long 
   wavelengths may also
   be caused by a very small, optically thick disk \citep[see, e.g.,][]{beckwith:1991}. To break this degeneracy it is necessary to resolve the disks to determine their 
   actual sizes, since if the physical disk size is known, a reasonable estimate of the disk's opacity can be obtained. To resolve the disks around T Tauri stars,
   interferometric observations are indispensable.
   
   Considerable progress has been made in this field over the last several years. \citet{wilner:2000} resolved the inner disk of the classical T Tauri star \object{TW Hya} in
   dust emission at 7 mm using the Very Large Array (VLA). \citet{calvet:2002} did extensive modelling of the SED of this source and showed that the dust grains in the disk must 
   have grown to sizes
   of at least $\sim 1$ cm. \citet{wilner:2003} used the Australia Telescope Compact Array (ATCA) to study TW Hya at 3.4~mm and found that 
   a passive two-layer disk \citep{chiang:1999} provides a reasonable model to explain the observations. Ten T Tauri stars in the Taurus-Auriga star-forming regions were
   resolved at 7 mm with the VLA by \citet{rodmann:2006}; the majority of these show strong evidence for grain growth to at least millimetre-sized dust.
   
   The more massive counterparts of the T Tauri stars, the Herbig Ae/Be stars, have been studied extensively. \citet{meeus:2001} classified 14 isolated Herbig Ae/Be stars
   into two groups based on the shape of the SED. 
   Disks with a flared
   outer part of the disk show a rising mid-infrared (20--100 $\mu$m) excess \citep[{\it Group I} in the classification by ][]{meeus:2001}, whereas self-shadowed disks have more
   modest mid-infrared excesses ({\it Group II}). \citet{acke:2004a} compared the millimetre slopes for a sample of 26 Herbig Ae/Be stars, based on single-dish data, and found that 
   {\it Group II} sources have in
   general a shallower slope, consistent with grain growth to larger sizes than in the {\it Group I} sources, and observationally confirming the model predictions from
   \citet{dullemond:2004}. \citet{natta:2004} 
   analysed interferometric observations of nine pre-main-sequence stars, mostly Herbig Ae stars, and find that the observations are well explained with dust size distributions 
   containing boulders of up to metre sizes.
   
   We have used the ATCA to observe 3.3-mm continuum emission from a sample of 15 southern T Tauri stars, five in Lupus and ten in Chamaeleon. Compared with the Taurus 
   star-forming cloud, the T Tauri stars in Lupus and Chamaeleon are generally somewhat older, and the Lupus clouds are much richer in very low mass stars than the 
   other clouds \citep{hughes:1994}. The sources IM Lup and WW Cha were also observed in spectral-line mode to search for HCO$^+$ as a tracer of the molecular gas. The 
   observations of the Lupus sources were followed up with 1.4-mm observations using the SubMillimeter Array (SMA). The sample and the
   observations are described in Sect.~\ref{sect: observations}. The basic results are presented in Sect.~\ref{sect: results} and further discussed in Sect.~\ref{sect: 
   discussion}. We will summarize our results and draw some conclusions in Sect.~\ref{sect: conclusions}.


\section{Observations}\label{sect: observations}
   
   \subsection{Source selection}\label{sect: source selection}
   
   The ATCA was used to observe 15 southern T Tauri stars (listed in Table~\ref{tab: source list}) at 3.3~mm. The sources were selected to overlap with the sample observed in the
   c2d programme. We furthermore selected sources with strong 1.3-mm fluxes \citep{henning:1993, nurnberger:1997} to improve the chances of detection at 3.3~mm.
   
   The distances used in this work are $150 \pm 20$~pc to Lupus I (HT~Lup, GW~Lup) and Lupus II (IM~Lup, RU~Lup), $200 \pm 20$~pc to Lupus III (HK~Lup) \citep[][]{comeron:2006}, 
   and $160 \pm 15$~pc to Chamaeleon I \citep{whittet:1997}. The distances to most of these pre-main-sequence stars are not very well constrained 
   \citep[see, e.g.,][for discussions on the distances to the Lupus clouds]{comeron:2006, vankempen:2006}. 
   The distance to the isolated source T~Cha, however, is known to be closer at $66^{+19}_{-12}$~pc \citep[Hipparcos,][]{vandenancker:1998}.
   
   \subsection{ATCA observations}\label{sect: atca observations}
   
   ATCA\footnote{The Australia Telescope Compact Array is part of the Australia Telescope which is funded by the Commonwealth of Australia for operation as a National 
   Facility managed by CSIRO} observations were carried out in July 2003, October 2004, 
   and August 2005. All sources were observed at 3.3-mm continuum in double sideband. The 
   primary beam of the ATCA antennae is $\approx 35 \arcsec$ at 3~mm. Each sideband consisted of 32 channels and had an effective total bandwidth of 128 MHz.
   Furthermore, the sources WW~Cha and IM~Lup were observed in a dual mode: the lower sideband was a 512-channel band with an effective total bandwidth of 8~MHz to provide 
   high-frequency resolution for detection of the HCO$^+$ $J$ = 1--0 line, and the upper sideband was used as a wideband channel with again 32 channels and a total bandwidth 
   of 128~MHz. These two sources were observed for a complete track to reach a similar RMS noise in the continuum as that for the other sources, and to maximize the
   possibility of detecting the HCO$^+$ $J$ = 1--0 line.
   The resulting velocity resolution of the narrow-band observations was 0.11 km~s$^{-1}$, the velocity coverage was $\sim$24 km~s$^{-1}$. 
   The ATCA was in the EW214 configuration at the time of the observations in 2003 (three antennae equipped with 3-mm receivers, baselines of 31--107 metres) and in the H214C 
   configuration in 2004 and 
   2005 (five antennae, baselines of 82--247 metres). 
   The data were calibrated and reduced with the MIRIAD package \citep{sault:1995}. The quasars \object{PKS 1622-297} and \object{PKS 1057-797} served as gain calibrators 
   for the Lupus and
   Chamaeleon sources, respectively, and the absolute fluxes were calibrated on \object{Mars} or \object{Uranus}. The calibration is estimated to have an uncertainty of 
   $\sim$20 \%. The 
   passbands were calibrated on the quasars \object{PKS 0537-441} and \object{PKS 1253-055}. 
   
\begin{table*}
 \caption[]{Source list of sources observed with the ATCA.}
 \label{tab: source list}
 \centering
  \begin{tabular}{llcclcll}
   \hline
   \hline
   Source                       	& Cloud		& Age$^\mathrm{a}$	& Spectral 		& Luminosity$^\mathrm{c}$ 	& Distance$^\mathrm{d}$	& When observed$^\mathrm{e}$ 		& Other names$^\mathrm{f}$				\\
                                	&		& (Myr)			& type$^\mathrm{b}$     & (L$_\odot$)			& (pc)  		& 	        			&							\\
   \hline
   \object{SY Cha}              	& Cha I		& 3--5			& M0       		& 0.35				& $160 \pm 15$		& 2005					& IRAS 10557-7655, SZ 3, \\
   					& 		&			&			&				&			&					& HBC 565, Cha T4		\\
   \object{CR Cha}              	& Cha I		& 1--2			& K0, K2       		& 2.8				& $160 \pm 15$		& 2003$^\dagger$; 2005			& IRAS 10578-7645, SZ 6, \\
   					&		&			&			&				&			&					& HBC 244, Cha T8		\\
   \object{CS Cha}              	& Cha I		& 2--3			& K4      		& 1.3				& $160 \pm 15$		& 2003$^\dagger$; 2005			& IRAS 11011-7717, SZ 9, \\
   					&		&			&			&				&			&					& HBC 569, Cha T11		\\
   \object{DI Cha}              	& Cha I		& 3--4			& G1, G2      		& 8.9				& $160 \pm 15$		& 2005            			& IRAS 11059-7721, SZ 19, \\
   					&		&			&			&				&			&					& HBC 245, Cha T26, KG 6	\\
   \object{KG 28}               	& Cha I		& ...			& K7			& 1.1      			& $160 \pm 15$		& 2005             			& SZ 22, Cha T29					\\
   \object{Glass I}$^\mathrm{g}$	& Cha I		& 2--3			& K4 			& 1.3				& $160 \pm 15$		& 2005					& IRAS 11068-7717, Cha T33, KG 39	\\
   \object{KG 49}               	& Cha I		& ...			& ...		     	& 15				& $160 \pm 15$		& 2005  				& IRAS 11072-7727, Cha IRN				\\
   \object{WW Cha}              	& Cha I		& 0.4--0.8		& K5       		& 2.2				& $160 \pm 15$		& 2003$^\dagger$; 2005  		& IRAS 11083-7618, SZ 34, \\
   					&		&			&			&				&			&					& HBC 580, Cha T44		\\
   \object{XX Cha}              	& Cha I		& 10--40		& M1       		& 0.10				& $160 \pm 15$		& 2005              			& IRAS 11101-7603, SZ 39, \\
   					&		&			&			&				&			&					& HBC 586, Cha T49		\\
   \object{T Cha}               	& Isolated	& $> 12.5$		& G2       		& 1.3				& $66^{+19}_{-12}$     	& 2003$^\dagger$; 2005              	& IRAS 11547-7904, HBC 591				\\
   \object{HT Lup}$^\mathrm{h}$		& Lup I		& 0.4--0.8		& K2       		& 6.0				& $150 \pm 20$		& 2003$^\dagger$; 2005			& IRAS 15420-3408, SZ 68, HBC 248			\\
   \object{GW Lup}              	& Lup I		& 1.3--3.2		& M2       		& 0.23				& $150 \pm 20$		& 2003$^\dagger$; 2005	        	& IRAS 15435-3421, SZ 71, HBC 249			\\
   \object{IM Lup}              	& Lup II	& 0.09--0.6		& M0       		& 1.3				& $150 \pm 20$		& 2003$^\dagger$; 2005     		& IRAS 15528-3747, SZ 82, HBC 605			\\
   \object{RU Lup}              	& Lup II	& 0.04--0.5		& K7-M0    		& 2.1				& $150 \pm 20$		& 2003$^\dagger$; 2005	        	& IRAS 15534-3740, SZ 83, HBC 251			\\
   \object{HK Lup}              	& Lup III	& 0.7--1.4		& M0       		& 0.62				& $200 \pm 20$		& 2003	        			& IRAS 16050-3857, SZ 98, HBC 616			\\
   \hline
  \end{tabular}
 \begin{list}{}{}
  \item[$^\mathrm{a}$] 	Ages adopted from \citet{hughes:1994}, \citet{lawson:1996}, and \citet{vandenancker:1998}.
  \item[$^\mathrm{b}$]	Spectral types adopted from \citet{herbig:1972}, \citet{gauvin:1992}, \citet{hughes:1994}, \citet{alcala:1995}, \citet{vandenancker:1998}, and 
  			\citet{comeron:1999}.
  \item[$^\mathrm{c}$]	Luminosities adopted from \citet{hughes:1994}, \citet{lawson:1996}, \citet{chen:1997}, and \citet{vandenancker:1998}.
  \item[$^\mathrm{d}$]	Distances adopted from \citet{whittet:1997}, \citet{vandenancker:1998}, \and \citet{comeron:2006}.
  \item[$^\mathrm{e}$]	The observations marked with a $\dagger$ were not used in the analysis.
  \item[$^\mathrm{f}$]  Catalogue names: KG = [KG2001] \citep{kenyon:2001}, IRAS = Infrared Astronomical Satellite, SZ = Southern Zwicky \citep{rodgers:1978}, HBC = 
 			Herbig+Bell Catalog \citep{herbig:1988}, Cha T = Assoc Chamaeleon T \citep{henize:1973}.
  \item[$^\mathrm{g}$]	This source is a binary with a separation of $2\farcs4$; the spectrum quoted is that of component A.
  \item[$^\mathrm{h}$]	This source is a binary in 2MASS K-band images with a separation $< 3\arcsec$. The spectrum quoted includes both sources.
 \end{list}
\end{table*}
		
   The phase centre was offset from the source position by $5 \arcsec$ to avoid 
   possible artefacts. There was some overlap in the samples that were observed in the three runs.
   The 2004 data suffered badly from unstable weather and were not used in the analysis,
   whereas the 2003 and 2005 data were consistent.
   In 2003, there were only three ATCA antennae fitted with 3-mm receivers, resulting in a large and elongated beam. Five antennae with 3-mm receivers were available in 
   2005, greatly
   improving the resolution and beam shape of our observations. With the exception of HK~Lup, which was only observed in 2003, we decided to use only the 2005 data for our
   analysis. For an overview of the ATCA observations, see Table~\ref{tab: observations}.
   
   \begin{table*}
   	\caption[]{Overview of the observations.}
   	\centering
   	\begin{tabular}{lccc}
		\hline
		\hline
		Obs. dates	& Freqs. covered	& ($u, v$) range covered	& \# Antennae used	\\
				& (GHz)			& (k$\lambda$)			&			\\
		\hline
		\multicolumn{4}{c}{ATCA}									\\
		\hline
		10 July 2003	& 89.999, 90.095	& 8-33				& 3			\\
		11 July 2003	& 89.999, 90.095	& 9-33				& 3			\\
		12 July 2003	& 89.999, 90.095	& 6-33				& 3			\\
		13 July 2003	& 89.999, 90.095	& 6-33				& 3			\\
		19 August 2005	& 89.181, 91.456	& 19-76				& 5			\\
		24 August 2005	& 89.180, 91.456	& 12-70				& 5			\\
				& 90.000, 90.096	& 14-75				& 5			\\
		25 August 2005	& 89.176, 91.456	& 10-71				& 5			\\
		26 August 2005	& 89.176, 91.456	& 12-71				& 5			\\
		28 August 2005	& 89.176, 91.456	& 12-71				& 5			\\
		\hline
		\multicolumn{4}{c}{SMA}										\\
		\hline
		28 April 2006	& 217.347, 226.892	& 4-53				& 8			\\
		\hline
	\end{tabular}
   \label{tab: observations}
   \end{table*}
   
   \subsection{SMA observations}
   
   The five Lupus sources in our sample were observed in double-sideband continuum (217 and 227 GHz, primary beam $\sim 55\arcsec$) with the SMA\footnote{The SubMillimeter 
   Array is a joint project between the Smithsonian Astrophysical Observatory and the Academia Sinica Institute of Astronomy and Astrophysics and is funded by the Smithsonian 
   Institution and the Academia Sinica.} \citep{ho:2004} on 28 April 2006. The data from 
   both sidebands were combined, giving an effective bandwidth of 3 GHz. The SMA was in the compact configuration and baselines ranged from 5 to 52 metres. The low elevation of 
   the sources as seen from Mauna Kea and the relatively short integration time of $\sim 30$ minutes per source resulted in a rather elongated beam of $\sim 9 \times 2.5$ arcsec 
   (natural weighting). The raw visibility data were calibrated and flagged with MIR, and 
   the calibrated visibility data were analysed with MIRIAD. The gains were calibrated on the quasar PKS~1622-297, and the absolute fluxes and correlator passbands were calibrated
   on Uranus with an expected uncertainty of $\sim$20 \%. 
   All eight antennae of the array were available at the time of the observations. For an overview of the SMA observations, see Table~\ref{tab: observations}.
   
   
\section{Results}\label{sect: results}

   \subsection{Source fluxes}\label{sect: source fluxes}
   
\begin{figure*}
 	\begin{flushleft}
  		\begin{center}
    			\vspace{-0.1cm}
    			\hspace{0cm}
    			\hbox{\psfig{figure=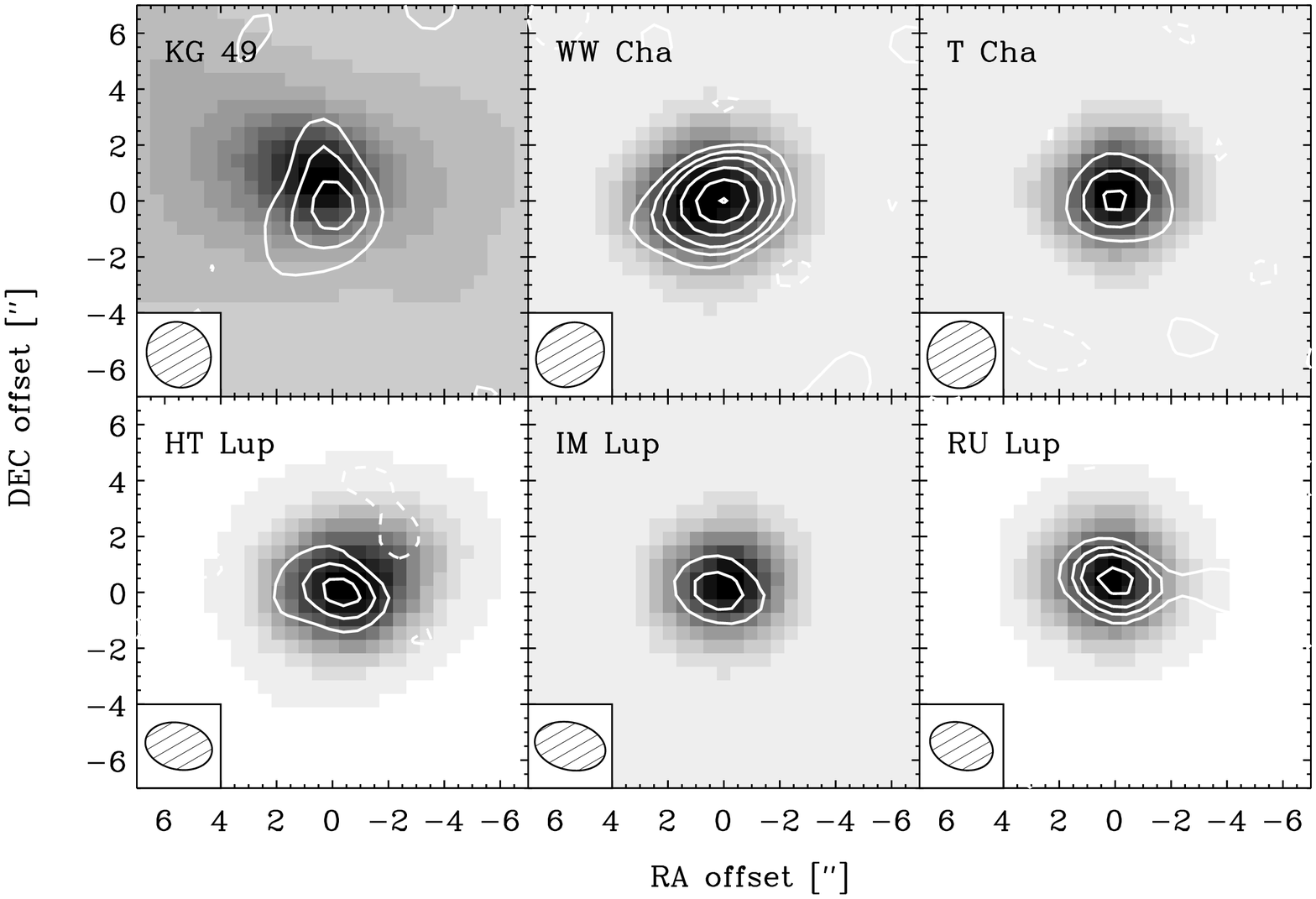,width=18cm}}
    			\hfill\parbox[b]{18cm}{\caption[]{ATCA images of the $\lambda = 3.3$-mm continuum emission (contours), overplotted on 2MASS K-band (2.1 $\mu$m) images 
			(grayscales). Contour levels are drawn at 2, 4, 6, 10, 15, and 20 times the noise level; negative contours are dashed. The positional offsets are with 
			respect to the fitted coordinates (see Table~\ref{tab: ATCA results}).}\label{fig: overlays}}
  		\end{center}
 	\end{flushleft}
   \end{figure*}

   From the 15 sources that were observed with the ATCA in 2003 and 2005, ten were detected at better than 3$\sigma$, with 3.3-mm fluxes ranging from $\sim$ 6~mJy up to $\sim$ 
   30~mJy. For those objects, both point sources and circular Gaussians were fitted in the ($u, v$) plane, the results of which are presented in Table~\ref{tab: ATCA results}. 
   Those sources that were detected at better than 5$\sigma$ are shown in Fig.~\ref{fig: overlays}, overplotted on 2MASS K-band (2.1~$\mu$m) images. The positions of the 
   infrared sources and the millimetre peaks agree very well.   

\begin{table*}
 \caption[]{Basic results of ATCA observations at 3.3~mm.}
 \label{tab: ATCA results}
 \centering
 \begin{tabular}{lcccccccc}
  \hline
  \hline
  Source name		& \multicolumn{2}{c}{Continuum flux$^\mathrm{a}$}	& RMS$^\mathrm{b}$	& Gaussian size		& RA$^\mathrm{a}$	& Dec.$^\mathrm{a}$	& Beam size$^\mathrm{c}$		& Effective frequency	\\
  			& \multicolumn{2}{c}{(mJy)}				& (mJy/beam)		& (arcsec)		& (J2000)		& (J2000)		& (arcsec)				& (GHz)			\\
			& (P)			& (G)				&			&			&			&			&					&			\\
  \hline
     SY Cha		& \multicolumn{2}{c}{$< 4.8^\mathrm{d}$}		& 1.6			& -			& 10 56 30.4		& -77 11 45.0		& $2.5 \times 2.2$			& 90.3			\\
     CR Cha		& $6.2$			& $6.1$				& 1.5			& $< 1$			& 10 59 06.9		& -77 01 39.7		& $2.5 \times 2.1$			& 90.3			\\
     CS Cha		& $5.9$			& $8.8$				& 1.5			& $1.65 \pm 0.96$	& 11 02 24.9		& -77 33 35.9		& $2.4 \times 2.2$			& 90.3			\\
     DI Cha		& \multicolumn{2}{c}{$< 3.9^\mathrm{d}$}		& 1.3			& - 			& 11 07 21.6		& -77 38 12.0		& $2.4 \times 2.2$			& 90.3			\\
     KG 28		& \multicolumn{2}{c}{$< 3.9^\mathrm{d}$}		& 1.3			& - 			& 11 07 57.9		& -77 38 50.0		& $2.4 \times 2.3$			& 90.3			\\
     Glass I		& \multicolumn{2}{c}{$< 3.0^\mathrm{d}$}		& 1.0			& - 			& 11 08 15.1		& -77 33 59.0		& $2.4 \times 2.4$			& 90.3			\\
     {\bf KG 49}	& $11.9$		& $29.5$			& 1.5			& $2.78 \pm 0.40$	& 11 08 38.6		& -77 43 52.1		& $2.4 \times 2.2$			& 90.3			\\
     {\bf WW Cha}	& $25.9$		& $33.1$			& 1.2			& $1.32 \pm 0.16$	& 11 10 00.0		& -76 34 58.0		& $2.5 \times 2.2$			& 91.0			\\
     XX Cha		& \multicolumn{2}{c}{$< 4.5^\mathrm{d}$}		& 1.5			& -			& 11 11 39.7		& -76 20 21.0		& $2.4 \times 2.2$			& 90.3			\\
     T Cha		& $6.4$			& $7.0$				& 1.0			& $0.78 \pm 1.03$	& 11 57 13.6		& -79 21 31.7		& $2.5 \times 2.4$			& 90.3			\\
     {\bf HT Lup}	& $8.3$			& $12.0$			& 1.1			& $1.40 \pm 0.41$	& 15 45 12.9		& -34 17 30.8		& $2.4 \times 1.7$			& 90.0			\\
     GW Lup		& $8.5$			& $9.5$				& 1.9			& $0.98 \pm 1.22$	& 15 46 44.7		& -34 30 36.0		& $5.3 \times 1.7$			& 90.0			\\
     {\bf IM Lup}	& $8.9$			& $13.1$			& 1.3			& $1.40 \pm 0.39$	& 15 56 09.2		& -37 56 06.3		& $2.3 \times 1.7$			& 91.5			\\
     {\bf RU Lup}	& $12.7$		& $15.6$			& 1.0			& $0.99 \pm 0.32$	& 15 56 42.3		& -37 49 16.0		& $2.3\times 1.7$			& 90.0			\\
     HK Lup		& $7.3$			& $9.1$				& 2.1			& $2.85 \pm 1.70$	& 16 08 22.5		& -39 04 46.3		& ill defined$^\mathrm{e}$		& 90.1			\\
  \hline
 \end{tabular}
\begin{list}{}{}
 \item[             ] {\sc Note.} --- Source name in boldface indicates that the source is resolved at 3.3~mm with the ATCA, see Sect.~\ref{sect: are the sources resolved?}.
 \item[$^\mathrm{a}$] Continuum flux and position are from fits in the ($u, v$) plane. For sources that were detected at 3$\sigma$, both the point-source flux (P)
 and the integrated flux for a Gaussian (G) are shown. For sources that were not 
 detected, the coordinates of the phase centre are quoted.
 \item[$^\mathrm{b}$] Calculated from the cleaned image.
 \item[$^\mathrm{c}$] Restored beam, using natural weighting.
 \item[$^\mathrm{d}$] Quoted value is 3$\sigma$ upper limit.
 \item[$^\mathrm{e}$] Large, elongated beam due to short integration time and availability of only three antennae with 3-mm receivers.
\end{list}
\end{table*}

	Table~\ref{tab: SMA results} presents the results of the SMA 1.4-mm observations of the Lupus sources, which were all detected at better than 15$\sigma$, with fluxes
	ranging from $\sim$ 70~mJy up to $\sim$ 210~mJy.
	For comparison, the SEST 1.3~mm single-dish fluxes from \citet{nurnberger:1997} are also shown. 
	
   \begin{table*}
   	\caption[]{Basic results of SMA observations at 1.4~mm. For comparison, the SEST 1.3~mm single-dish fluxes from \citet{nurnberger:1997} are shown.}
	\label{tab: SMA results}
	\centering
	\begin{tabular}{lccccccccc}
		\hline
		\hline
		Source name	& \multicolumn{2}{c}{Continuum flux$^\mathrm{a}$}	& RMS$^\mathrm{b}$	& Gaussian size		& RA$^\mathrm{a}$	& Dec.$^\mathrm{a}$	& Effective frequency  & \multicolumn{2}{c}{SEST 1.3 mm$^\mathrm{c}$}	  \\
				& \multicolumn{2}{c}{(mJy)}				& (mJy/beam)		& (arcsec)		& (J2000)		& (J2000)		& (GHz) 	       & Flux	  & RMS 				  \\
				& (P)			& (G)				&			&			&			&			&		       & (mJy)    & (mJy)				  \\
		\hline
		HT Lup		& 73			& 77				& 4.0			& $1.01 \pm 0.66$	& 15 45 12.9		& -34 17 30.1		& 221.3		       & 135	  & 15  				  \\
		GW Lup		& 64 			& 70				& 3.7			& $1.22 \pm 0.48$	& 15 46 44.8		& -34 30 35.7		& 221.3		       & 106	  & 18  				  \\
		{\bf IM Lup}	& 188			& 214				& 4.3			& $1.33 \pm 0.20$	& 15 56 09.2		& -37 56 06.5		& 221.3		       & 260	  & 9					  \\
		{\bf RU Lup}	& 148			& 158				& 4.5			& $1.02 \pm 0.32$	& 15 56 42.3		& -37 49 15.9		& 221.3		       & 197	  & 7					  \\
		{\bf HK Lup}	& 89			& 101				& 3.9			& $1.43 \pm 0.38$	& 16 08 22.5		& -39 04 47.5		& 221.3		       & 84	  & 17  				  \\
		\hline
	\end{tabular}
	\begin{list}{}{}
 		\item[             ] {\sc Note.} --- Source name in boldface indicates that the source is resolved at 1.4~mm with the SMA, see Sect.~\ref{sect: are the sources 
		resolved?}.
		\item[$^\mathrm{a}$] Continuum flux and position are taken from fits in the ($u, v$) plane. Both the point-source flux (P) and the integrated flux
		for a Gaussian (G) are shown.
		\item[$^\mathrm{b}$] Calculated from the cleaned image.
 		\item[$^\mathrm{c}$] \citet{nurnberger:1997}. The SEST fluxes are in general higher than the SMA fluxes, partly due to the slightly shorter
		effective wavelength, 1.3~mm (SEST) vs. 1.4~mm (SMA).
	\end{list}
   \end{table*}

   \subsection{Are the sources resolved?}\label{sect: are the sources resolved?}
   
\begin{figure*}
   	\begin{flushleft}
		\begin{center}
			\vspace{-0.1cm}
			\hbox{\hspace{-0.5cm}\psfig{figure=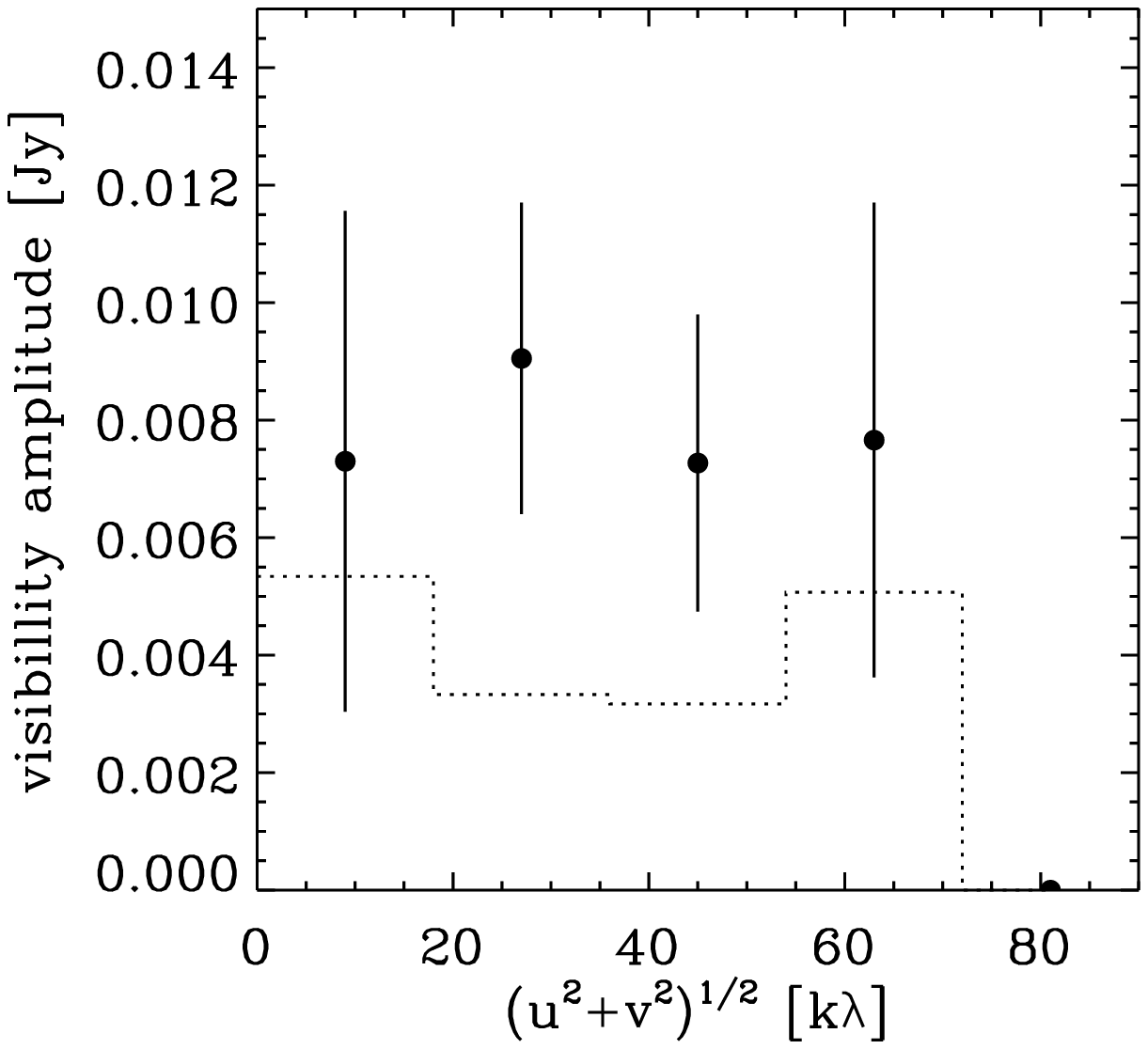,width=7.5cm}
			\put(-75,130){{\bf CR Cha}}
			\hspace{-1.5cm}\psfig{figure=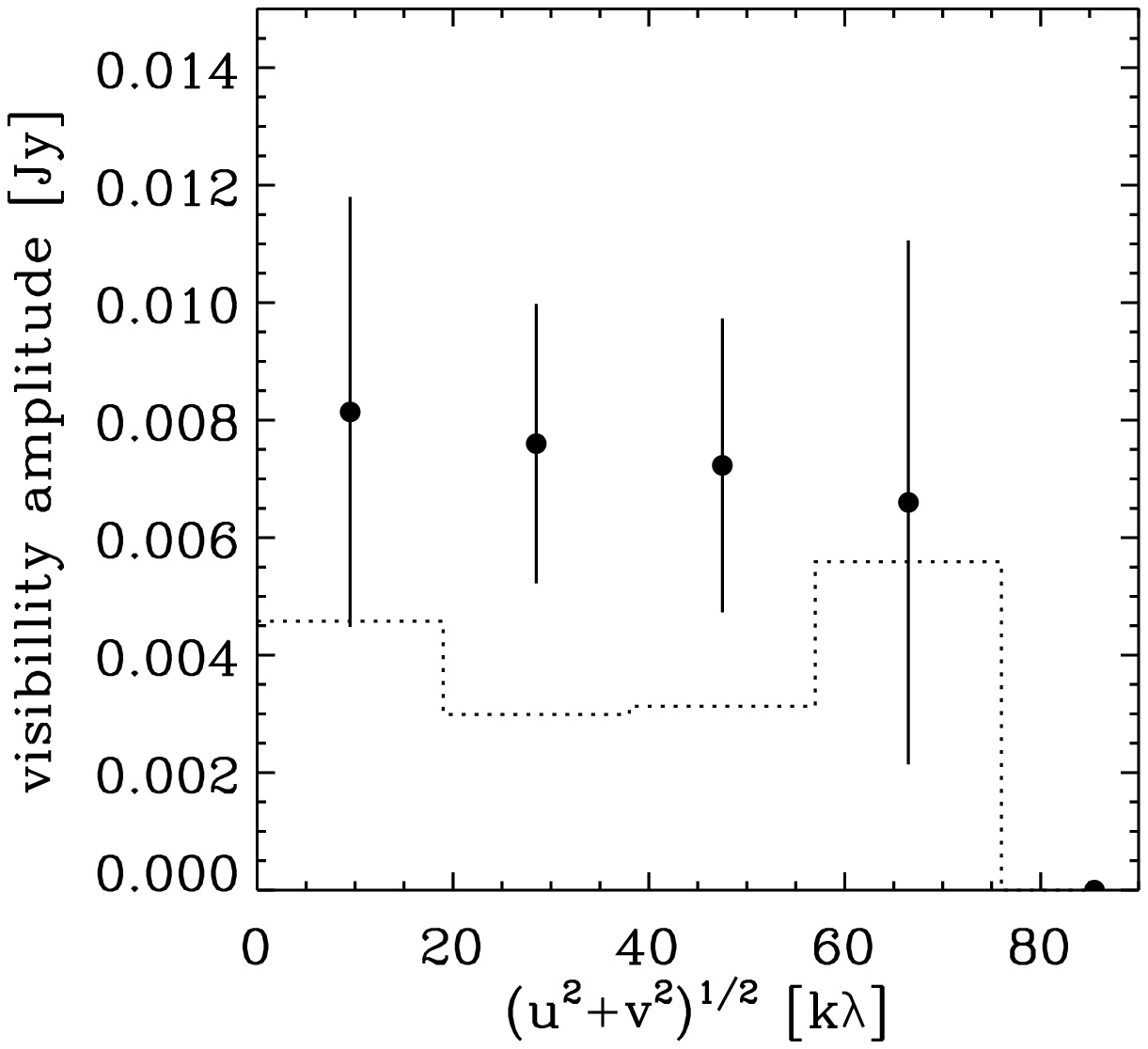,width=7.5cm}
			\put(-75,130){{\bf CS Cha}}
			\hspace{-1.5cm}\psfig{figure=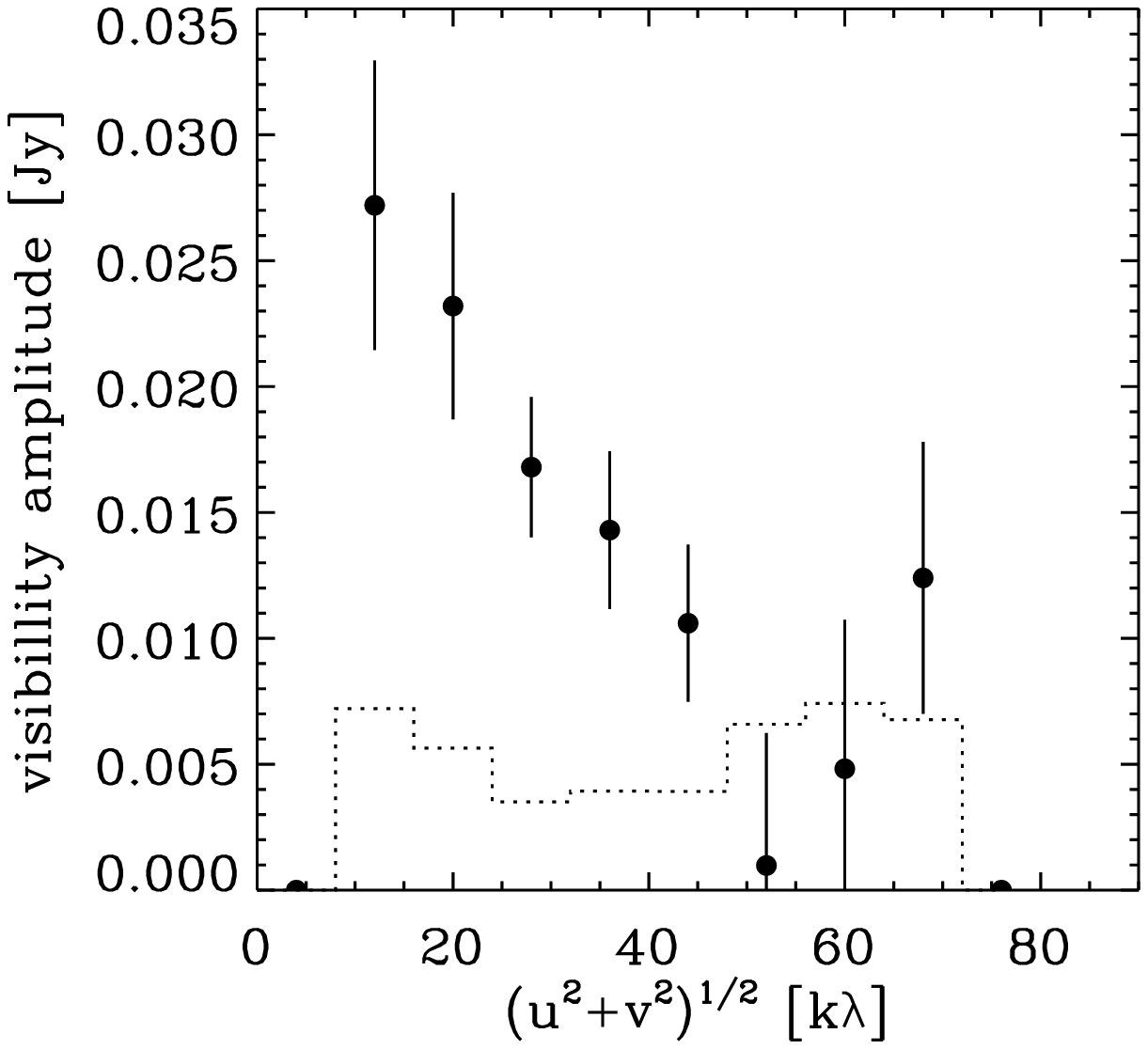,width=7.5cm}
			\put(-75,130){{\bf KG 49}}}
			\hbox{\hspace{-0.5cm}\psfig{figure=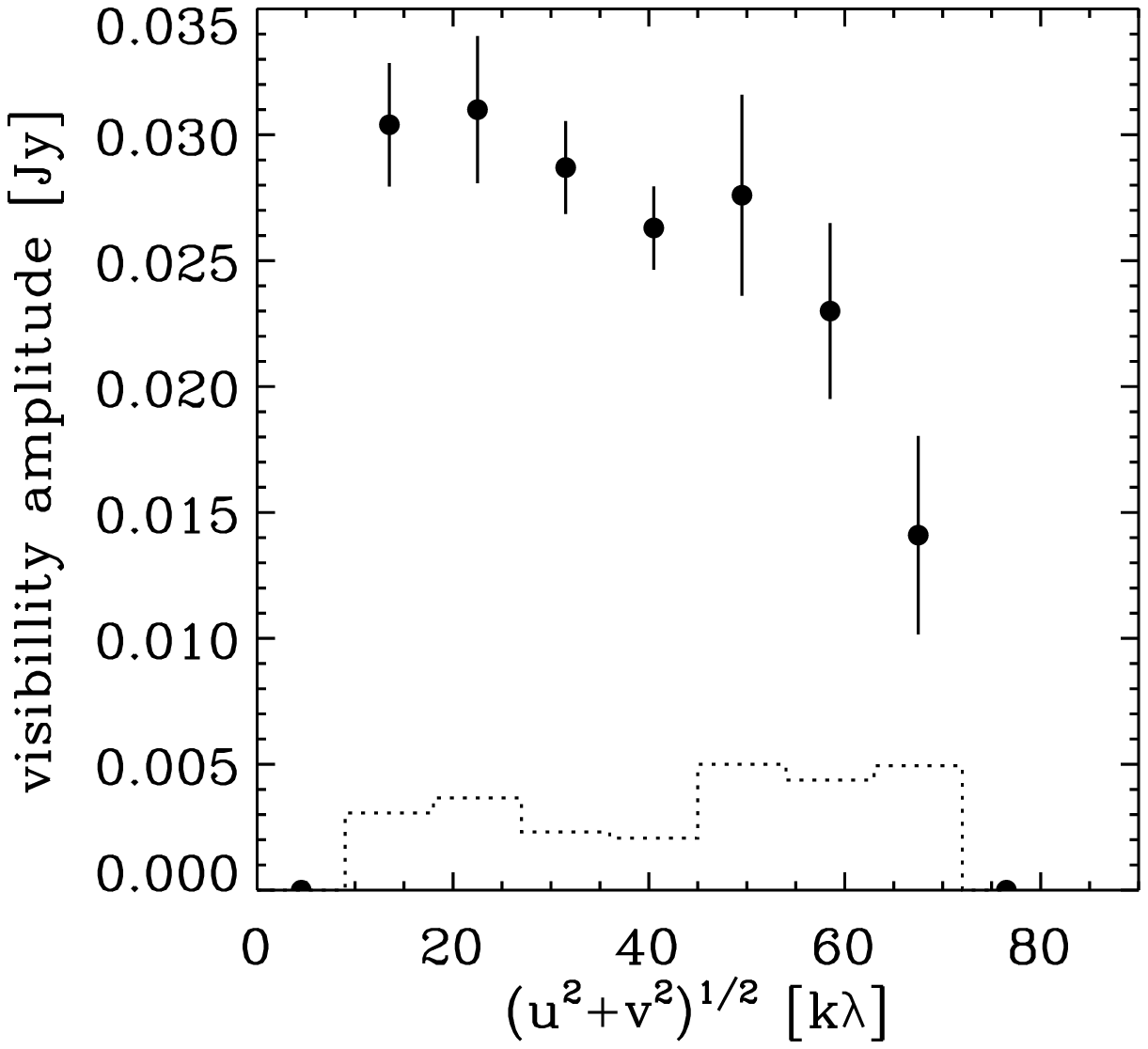,width=7.5cm}
			\put(-75,130){{\bf WW Cha}}
			\hspace{-1.5cm}\psfig{figure=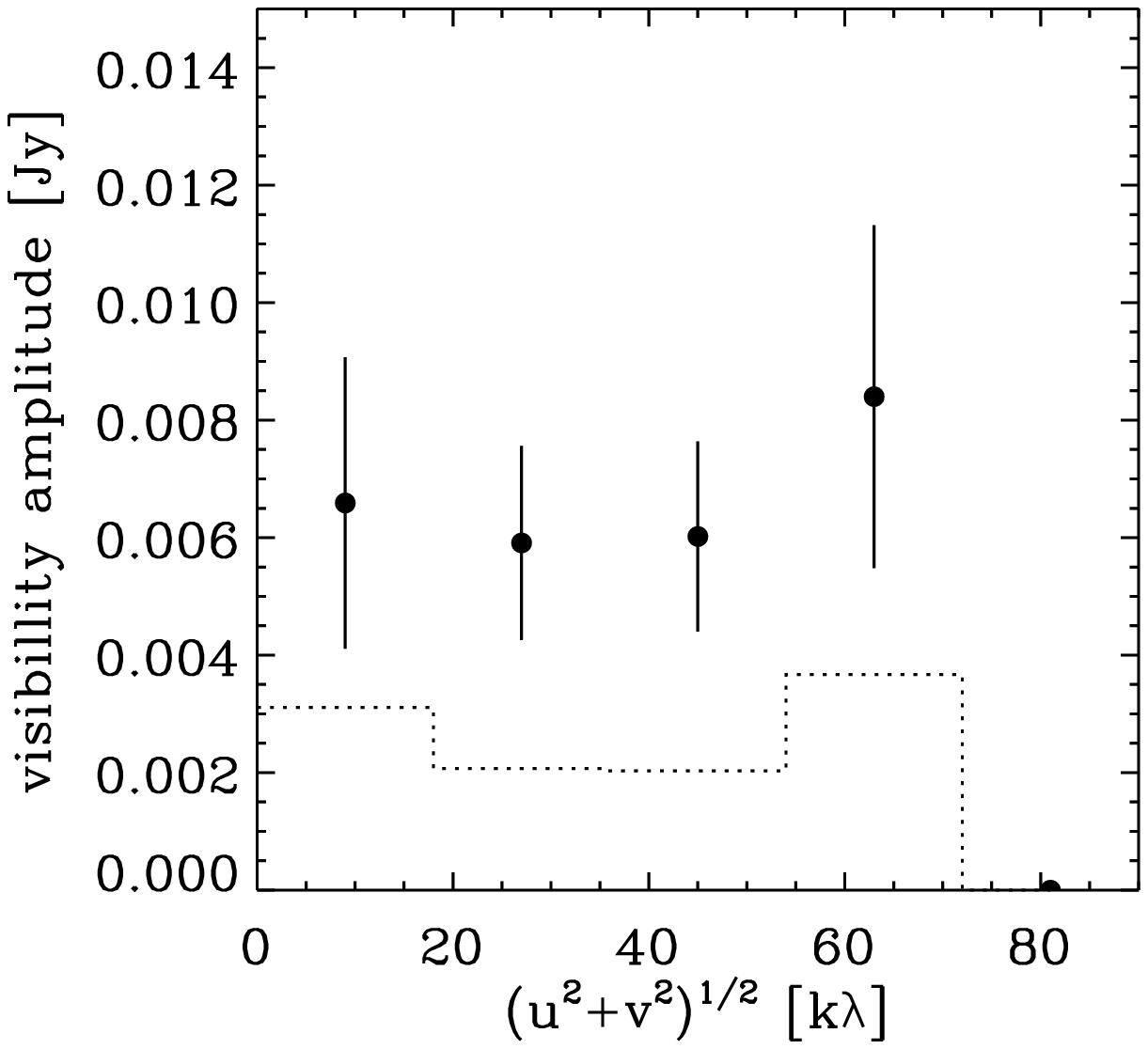,width=7.5cm}
			\put(-75,130){{\bf T Cha}}
			\hspace{5.6cm}}
			\hbox{\hspace{-0.5cm}\psfig{figure=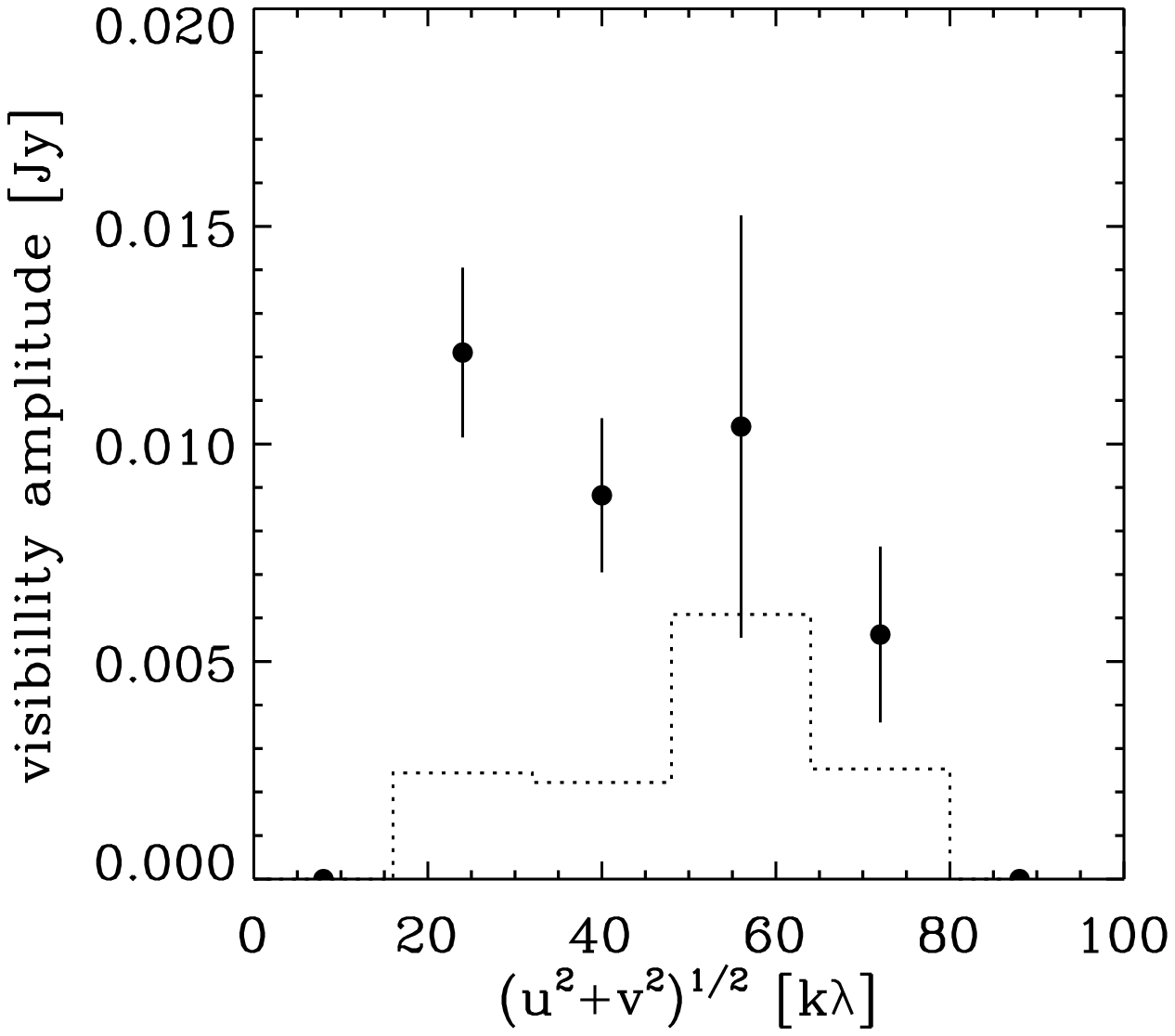,width=7.5cm}
			\put(-75,130){{\bf HT Lup}}
			\hspace{-1.5cm}\psfig{figure=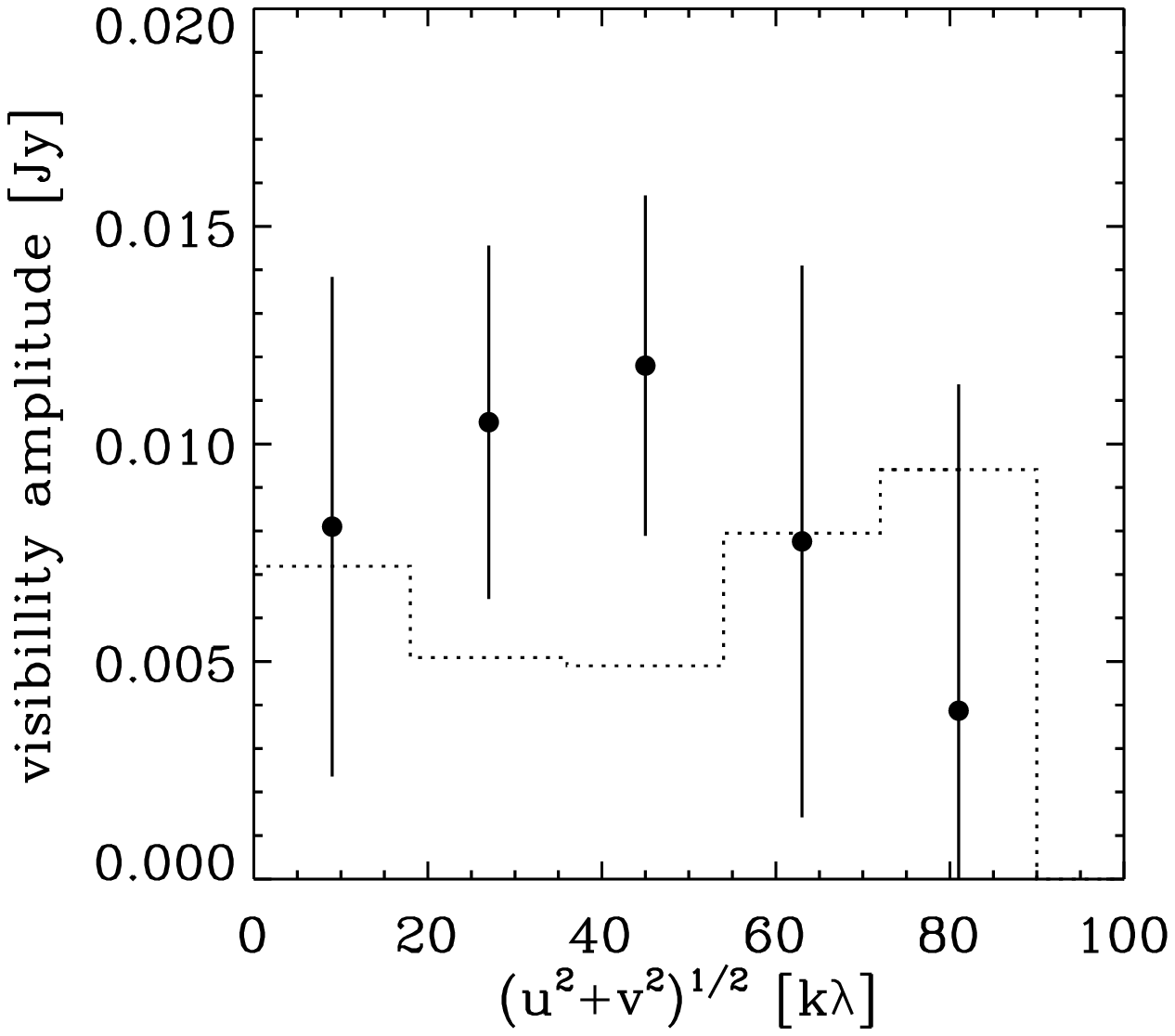,width=7.5cm}
			\put(-75,130){{\bf GW Lup}}
			\hspace{-1.5cm}\psfig{figure=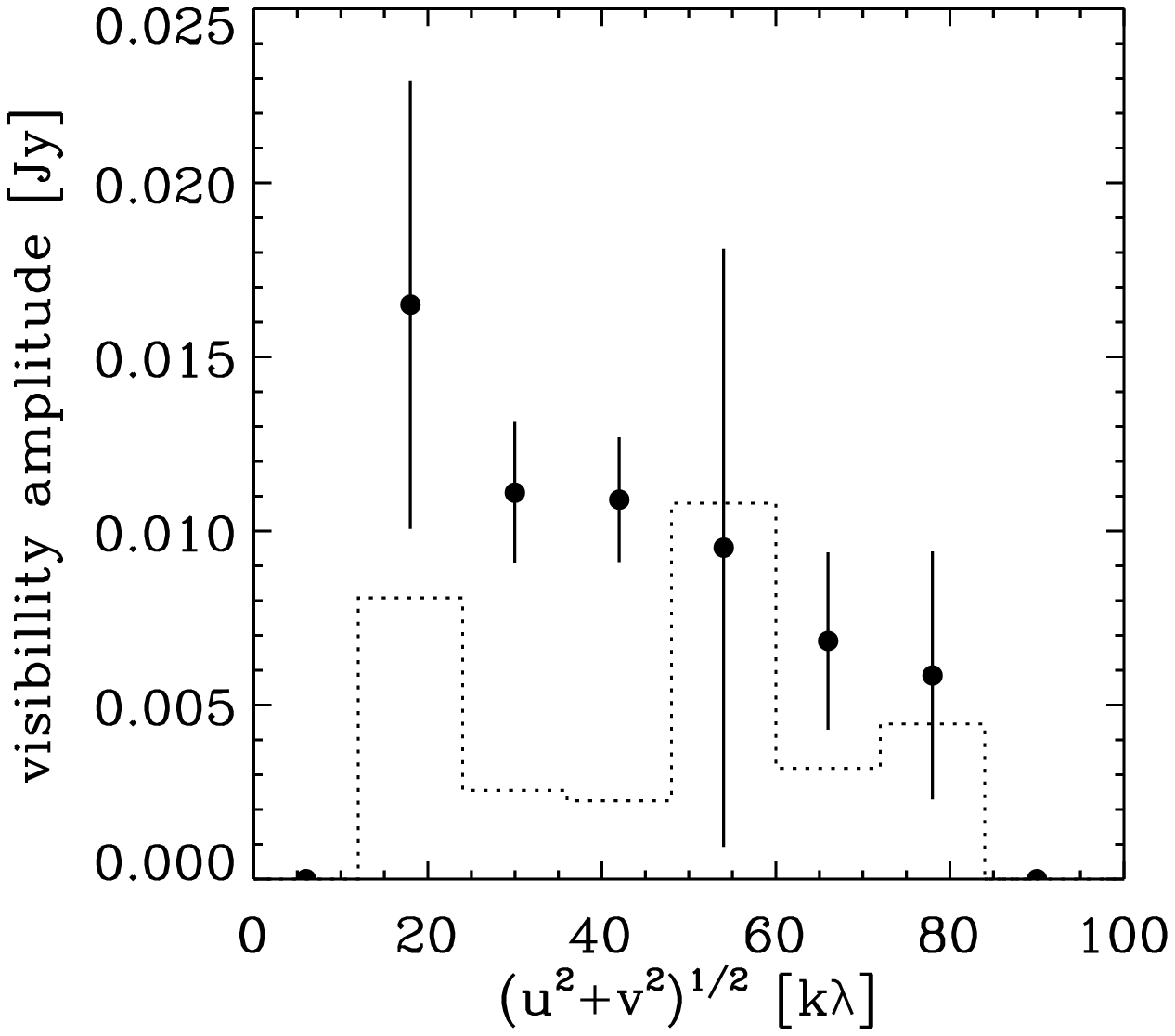,width=7.5cm}
			\put(-75,130){{\bf IM Lup}}}
			\hbox{\hspace{-0.5cm}\psfig{figure=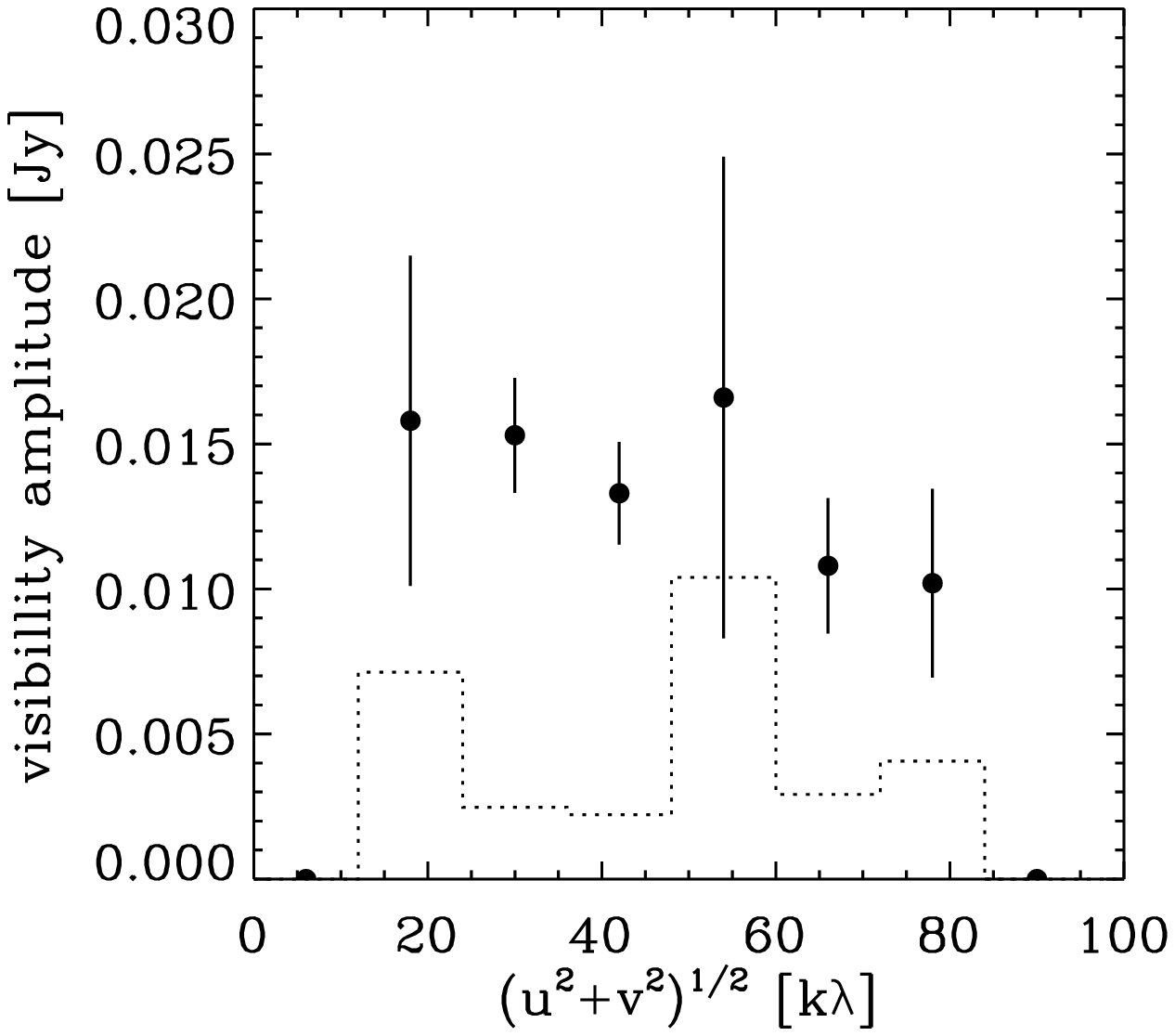,width=7.5cm}
			\put(-75,130){{\bf RU Lup}}
			\hspace{-1.5cm}\psfig{figure=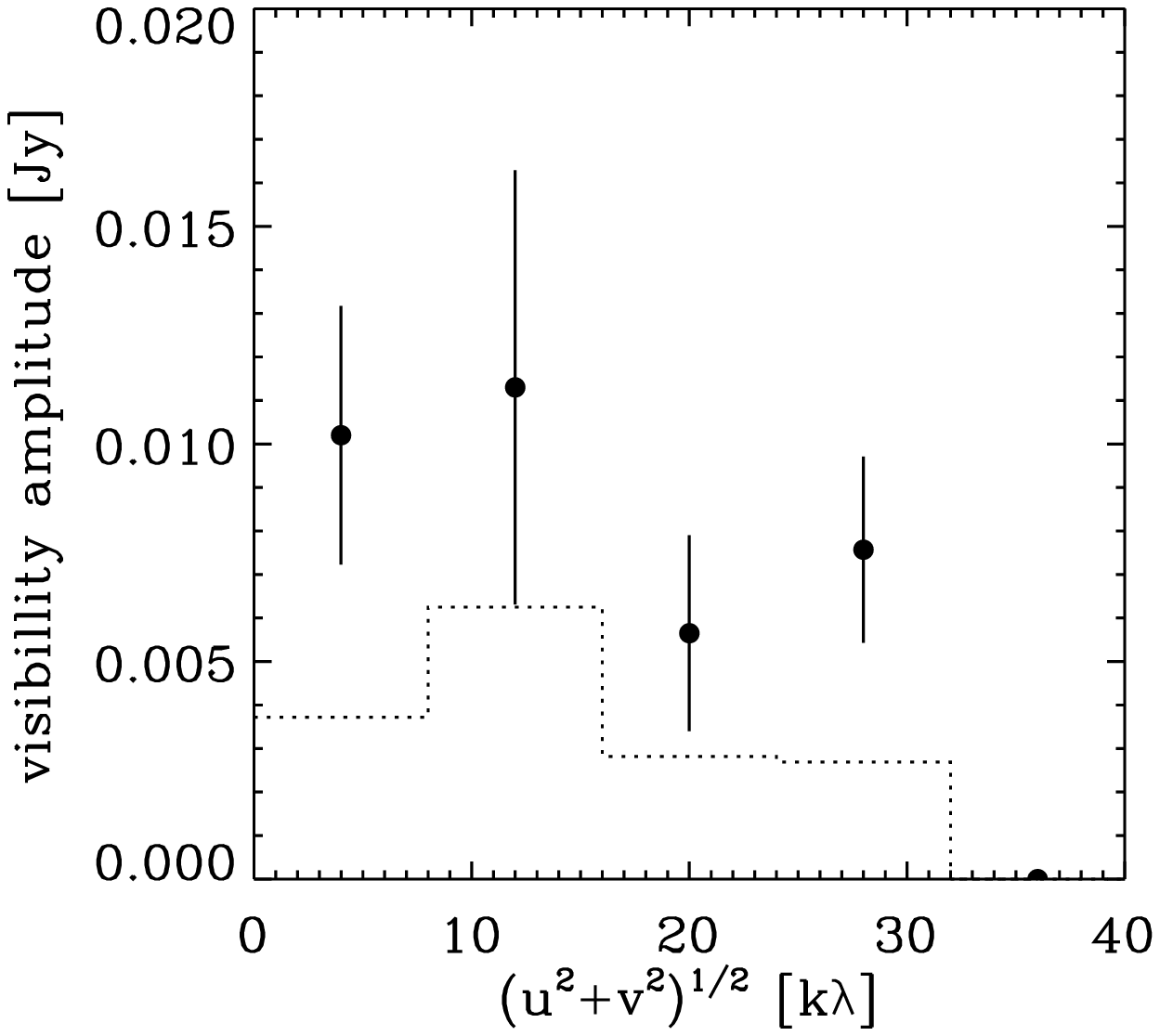,width=7.5cm}
			\put(-75,130){{\bf HK Lup}}
			\hspace{5.6cm}}
			\hfill\parbox[b]{18cm}{\caption[]{Amplitude vs. ($u, v$) distance for sources observed with the ATCA. The data points give the vector-averaged amplitude per
			bin, where the data are binned in annuli according to ($u, v$) distance. The error bars show the statistical $1\sigma$ errors, and the dotted lines give 
			the expected amplitude for zero signal.}\label{fig: UVamp atca}}
		\end{center}
	\end{flushleft}
   \end{figure*}

\begin{figure*}
 \begin{flushleft}
  \begin{center}
    	\vspace{-0.1cm}
    	\hspace{0cm}
    	\hbox{\hspace{-0.5cm}\psfig{figure=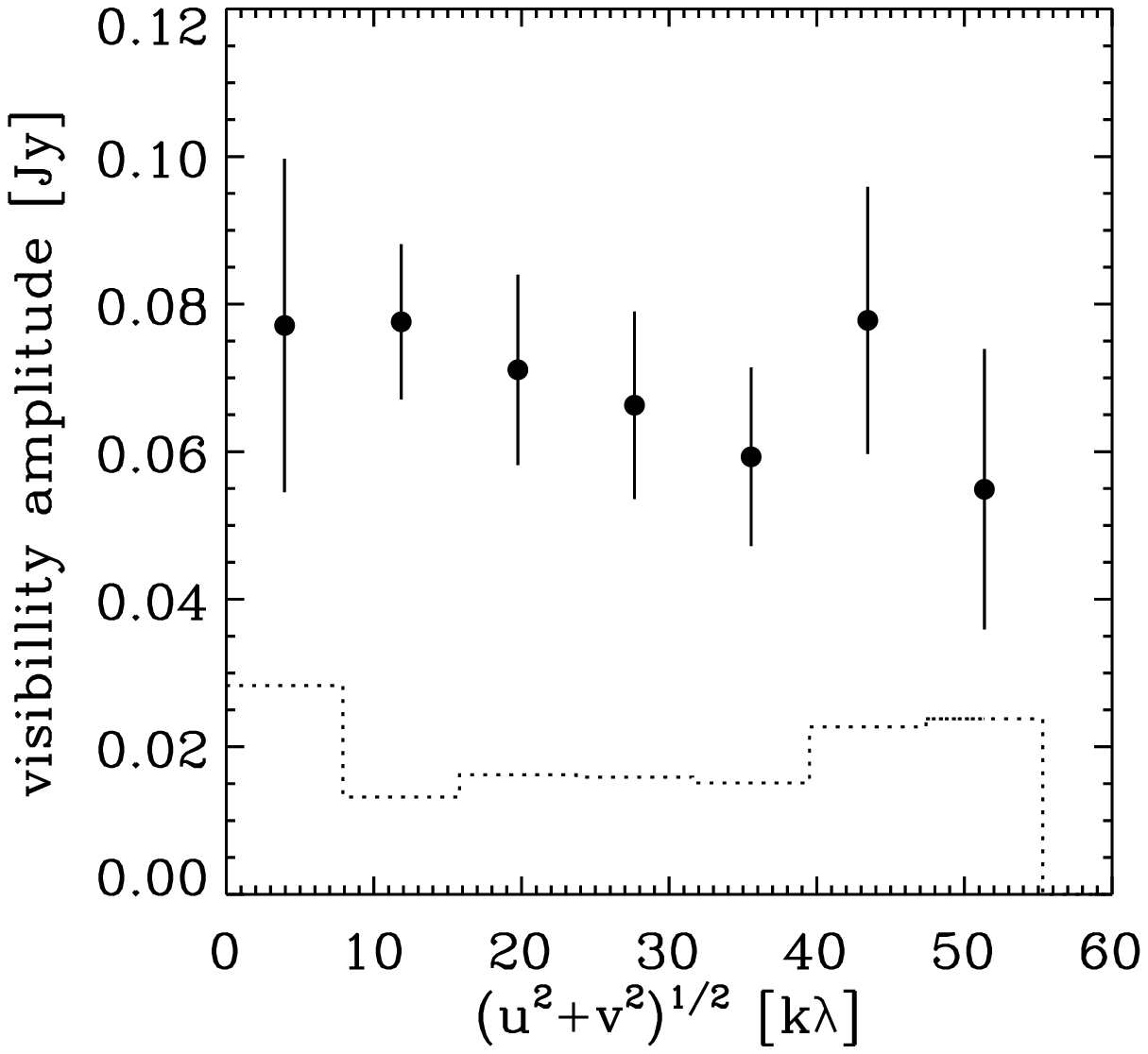,width=7.5cm}
    	\put(-75,130){{\bf HT Lup}}
    	\hspace{-1.5cm}\psfig{figure=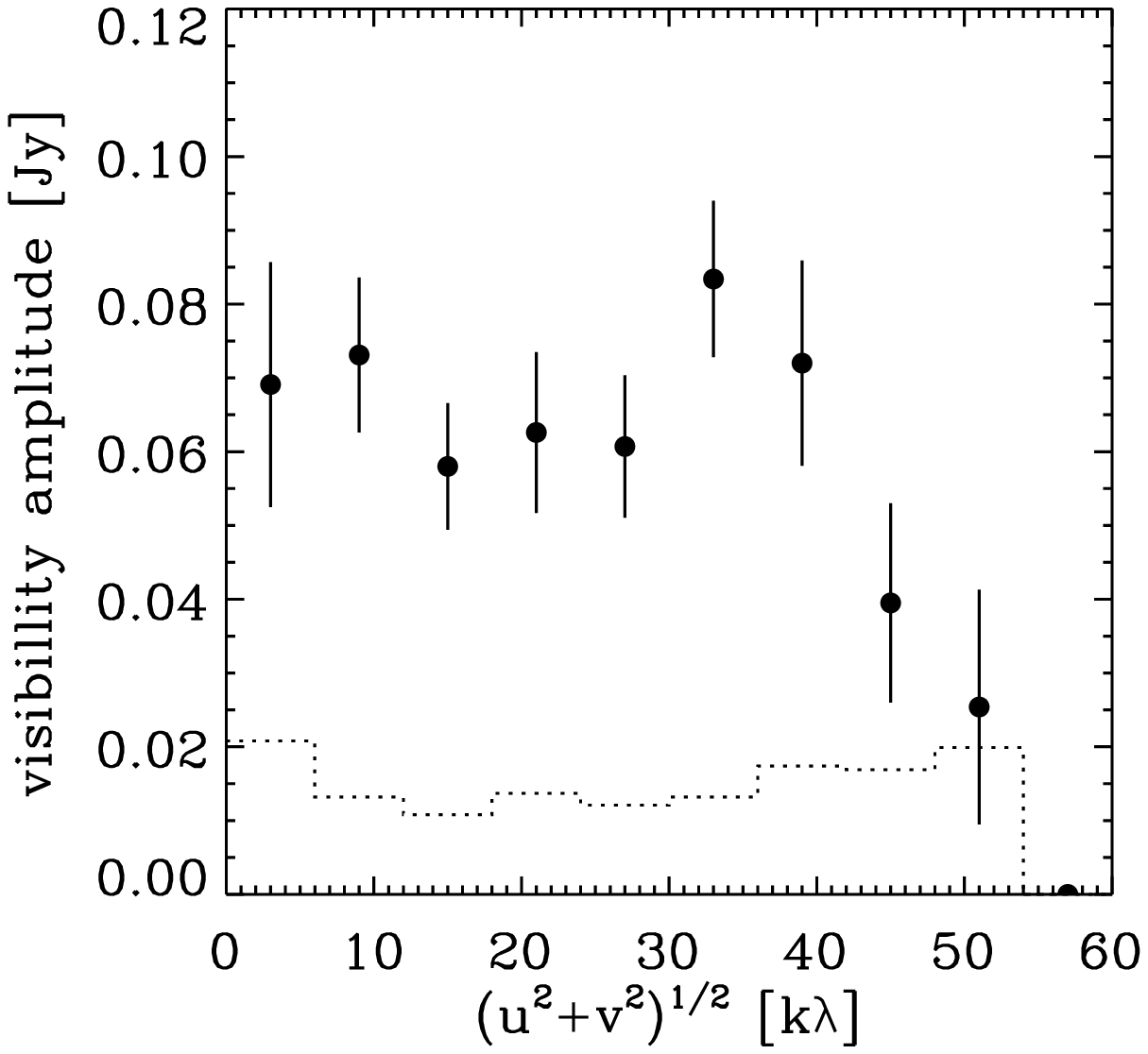,width=7.5cm}
    	\put(-75,130){{\bf GW Lup}}
    	\hspace{-1.5cm}\psfig{figure=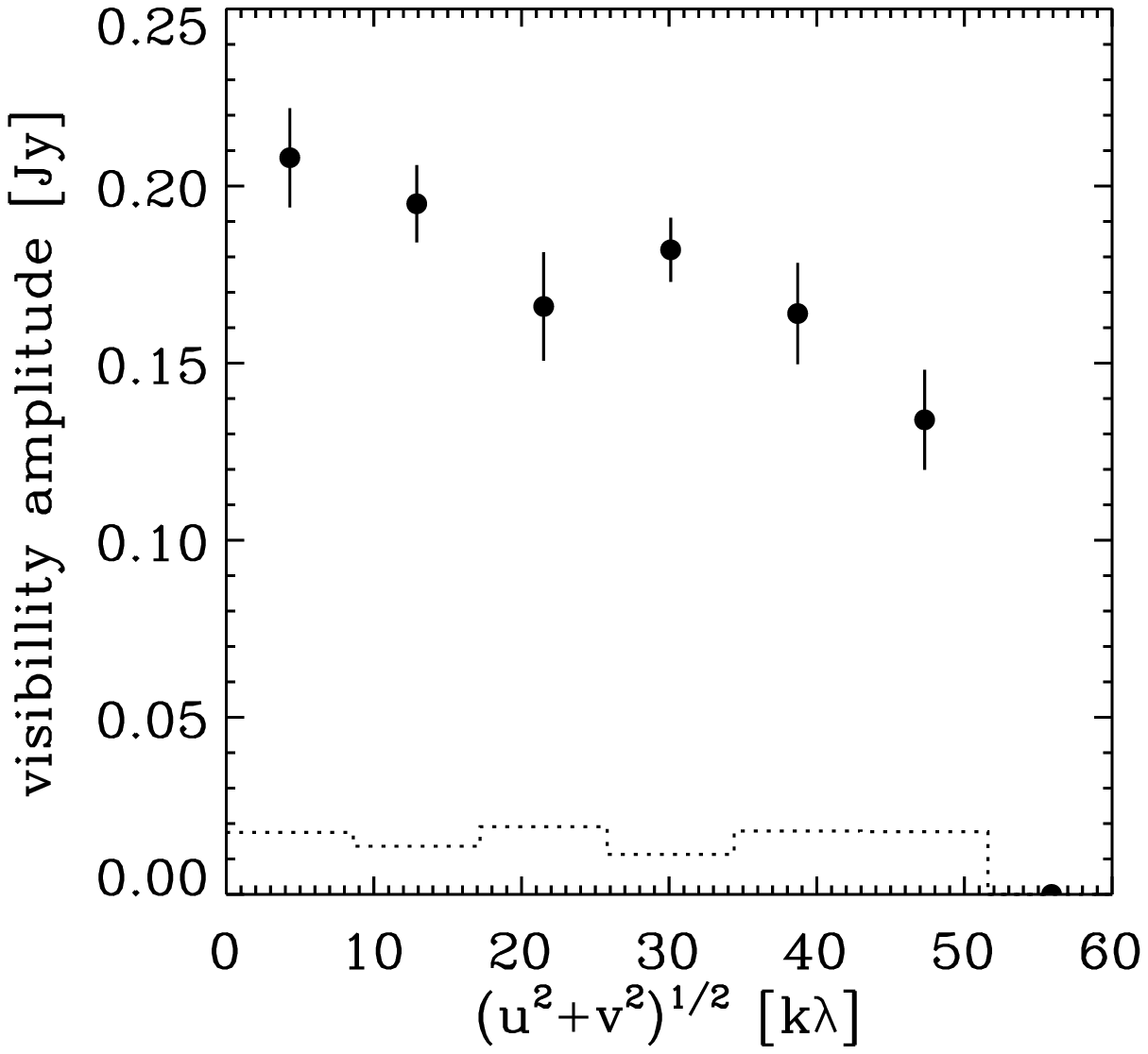,width=7.5cm}
    	\put(-75,130){{\bf IM Lup}}}
    	\hbox{\hspace{-0.5cm}\psfig{figure=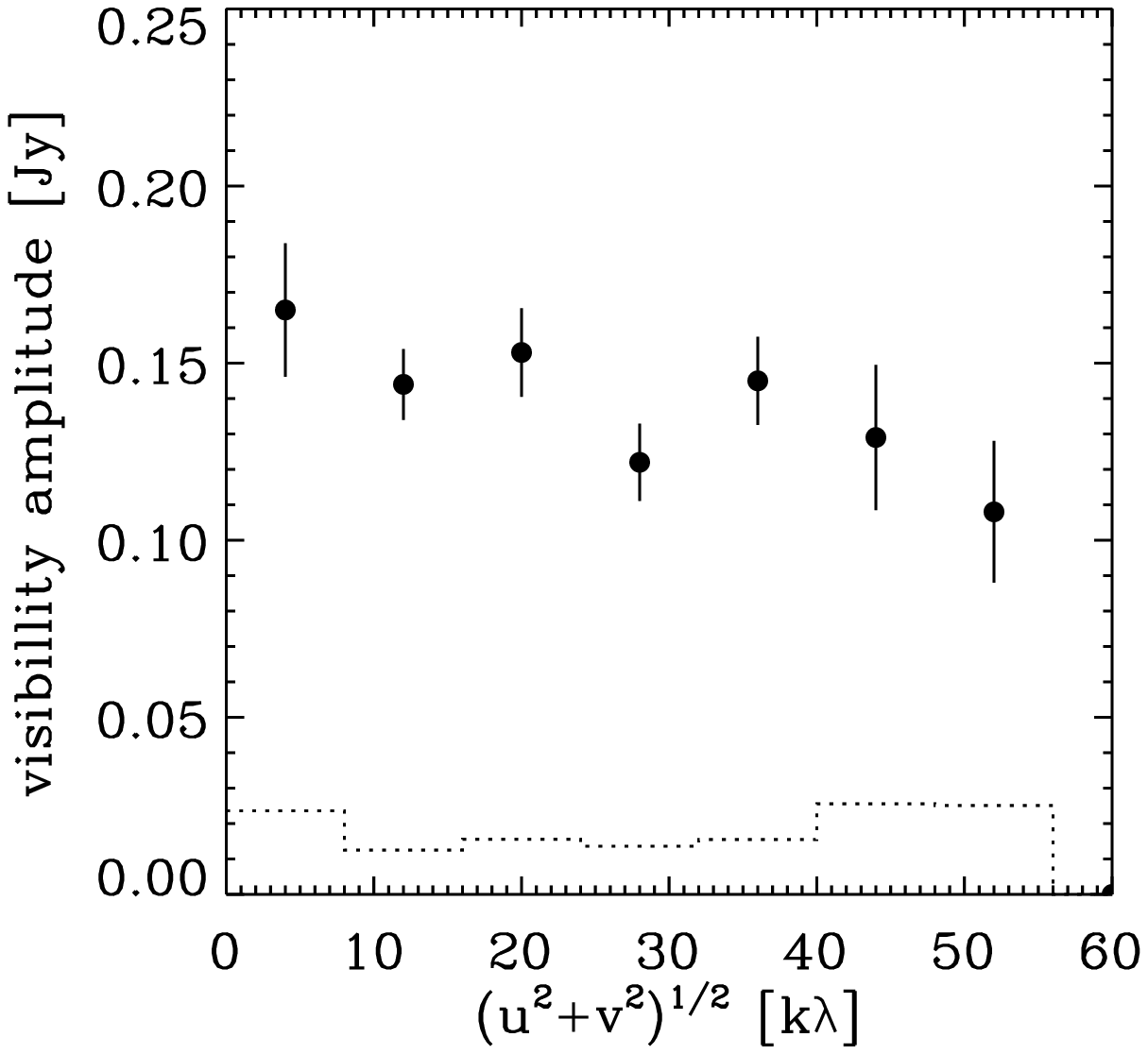,width=7.5cm}
    	\put(-75,130){{\bf RU Lup}}
    	\hspace{-1.5cm}\psfig{figure=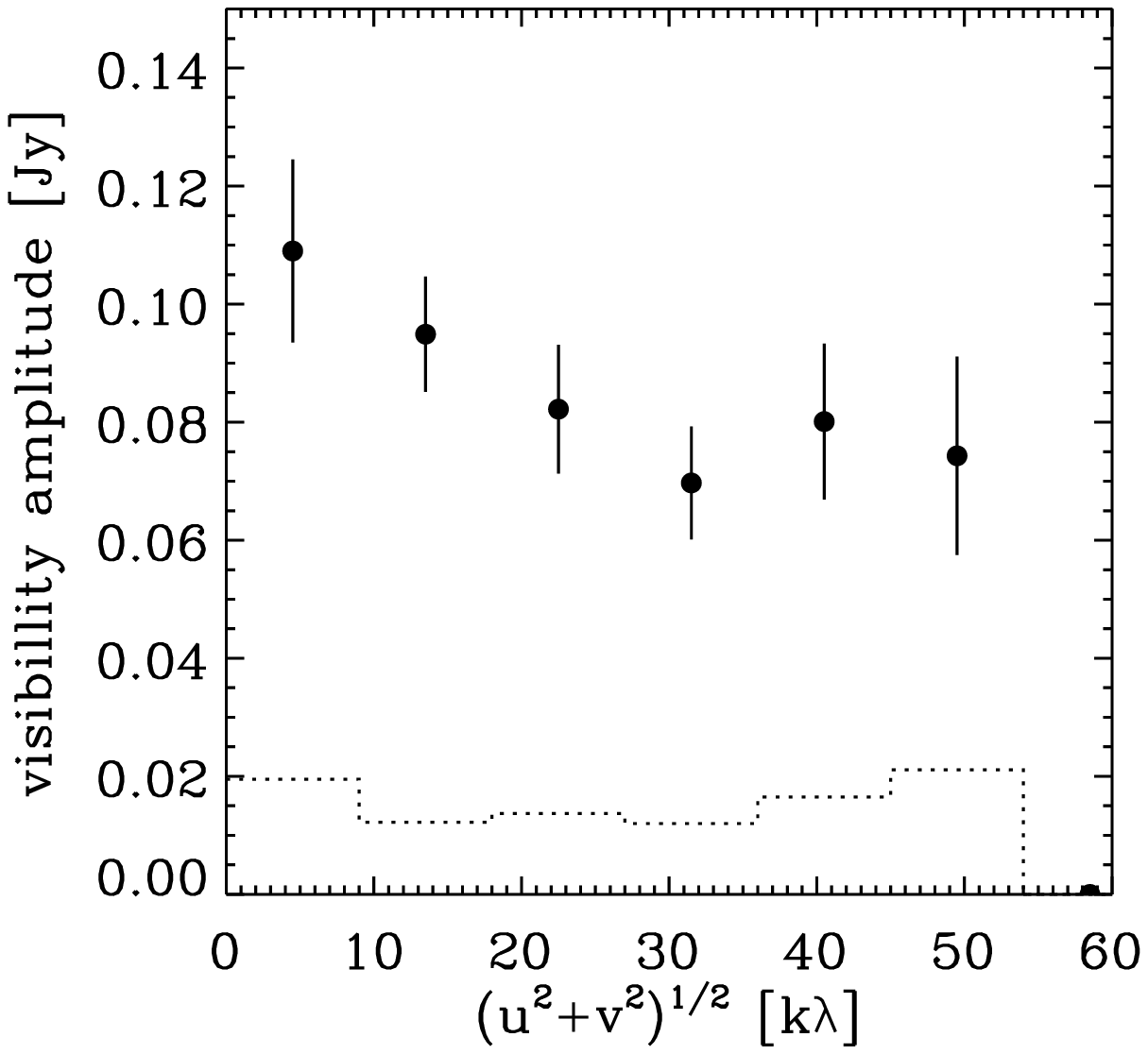,width=7.5cm}
    	\put(-75,130){{\bf HK Lup}}
	\hspace{5.6cm}}
  	\hfill\parbox[b]{18cm}{\caption[]{Same as Fig.~\ref{fig: UVamp atca} for sources observed with the SMA.}\label{fig: UVamp sma}}
  \end{center}
 \end{flushleft}
\end{figure*}

	A plot of the visibility amplitude as a function of baseline for each of our target sources is presented in Figs.~\ref{fig: UVamp atca} and \ref{fig: UVamp sma}. An 
	amplitude that decreases as a function of baseline indicates a resolved source, suggesting that at least some of our sources are resolved. Note that in principle 
	the resolved structure could be the protoplanetary disk, or the remnant centrally-condensed envelope around it. One can distinguish between the two by looking in more 
	detail at the shape of the amplitude vs. ($u, v$) distance curves. 
	A disk shows a shallower profile at shorter baselines, levelling off to the flux of the integrated disk emission. A power-law envelope shows a flux that 
	increases steeply towards the shortest baselines, and the total integrated flux can only be obtained with single-dish telescopes, observing with a larger beam. 
	An illustrative example is provided by Fig.~3 in \citet{jorgensen:2005}, who need a disk and an envelope to 
	explain the amplitude as a function of baseline for the deeply embedded class 0 source \object{NGC 1333 IRAS 2A}.
	 
	The amplitude vs. ($u, v$) distance plot for KG~49 (Cha~IRN) appears to indicate a resolved envelope rather than a disk. This is in line with the results of 
	\citet{henning:1993} who find that the SED from 1~$\mu$m to 1~mm cannot be fitted with a disk, but is consistent with a spherically symmetric model with constant 
	density and $A_V=25$ mag. Furthermore, all other sources show silicates in emission around 10~$\mu$m, which is indicative of a disk without significant foreground 
	absorption. KG~49 on the other hand does not show silicate emission around 10~$\mu$m. However, it does show bands of H$_2$O, CO, and CO$_2$ ice 
	\citep{gurtler:1999, pontoppidan:2003}, which can be explained by a cold envelope. 
	It thus appears that KG~49 is not a genuine T Tauri star, but a less evolved object that is still embedded in an envelope. We will therefore disregard KG~49 in the 
	further analysis in this work.
	
	We consider a source to be resolved if the integrated flux of the fitted Gaussian is at least 2$\sigma$ higher than the flux obtained from a point-source fit (see 
	Tables~\ref{tab: ATCA results} and \ref{tab: SMA results}). Note that for a point source the peak flux density equals the integrated flux. 
	According to this definition, five of the detected sources (KG~49, WW~Cha, HT~Lup, IM~Lup, RU~Lup) are resolved by the ATCA, and two of the Lupus sources (IM~Lup, HK~Lup) 
	are resolved by the SMA. In the case of KG~49 it is the envelope that is resolved, in the other cases the disks.

   \begin{table}
   	\caption[]{SEST 1.3~mm fluxes from the literature, mm slopes, derived opacity indices, and disk masses.}
	\label{tab: Gaussian results}
	\centering
	\begin{tabular}{lcccc}
		\hline
		\hline
		Source   		& SEST 1.3~mm$^\mathrm{a}$	& Mm  				& Opacity			& Disk mass$^\mathrm{c}$	\\
		name	     		& Flux				& slope$^\mathrm{b}$  		& index$^\mathrm{b}$   		& $M_{\rm disk}$		\\
			     		& (mJy)				& $\alpha$		     	& $\beta$	    		& (M$_\odot$)		     	\\
		\hline	
		SY Cha			& $< 172$			& ...				& ...				& $< 0.011$			\\
		CR Cha	     		& $125 \pm 24$			& $3.2 \pm 0.5$	     		& $1.5 \pm 0.6$			& $0.014$	     		\\
		CS Cha	     		& $128 \pm 46$			& $2.9 \pm 0.5$	     		& $1.0 \pm 0.6$			& $0.021$	     		\\
		DI Cha			& $38 \pm 11$			& $> 2.4$			& $> 0.5$			& $< 0.009$			\\
		KG 28			& ...				& ...				& ...				& $< 0.009$			\\
		Glass I			& $70 \pm 22$			& $> 3.4$			& $> 1.7$			& $< 0.007$			\\
		WW Cha	     		& $408 \pm 29$			& $2.7 \pm 0.7$	     		& $0.8 \pm 0.8$			& $0.077$	     		\\
		XX Cha			& $< 252$			& ...				& ...				& $< 0.011$			\\
		T Cha	     		& $105 \pm 18$			& $2.9 \pm 0.5$	     		& $1.1 \pm 0.6$			& $0.003$		     	\\
		HT Lup	     		& $135 \pm 15$			& $2.5 \pm 0.4$	     		& $0.4 \pm 0.5$			& $0.025$		     	\\
		GW Lup	     		& $106 \pm 18$			& $2.4 \pm 0.4$	     		& $0.5 \pm 0.5$			& $0.019$		     	\\
		IM Lup	     		& $260 \pm 9$			& $3.2 \pm 0.5$	     		& $1.4 \pm 0.5$			& $0.027$		     	\\
		RU Lup$^\mathrm{c}$	& $197 \pm 7$			& $2.5 \pm 0.1$	     		& $0.5 \pm 0.1$ 		& $0.032$		     	\\
		HK Lup	     		& $84 \pm 17$			& $2.5 \pm 0.4$	     		& $0.7 \pm 0.5$			& $0.033$		     	\\
		\hline
	\end{tabular}
	\begin{list}{}{}
		\item[$^\mathrm{a}$] \citet{henning:1993, nurnberger:1997}.
		\item[$^\mathrm{b}$] Calculated from SMA and ATCA fluxes (Gaussian fits) from this work and SEST fluxes, where available.
		\item[$^\mathrm{c}$] Disk masses estimated from ATCA fluxes (Gaussian fits), assuming a gas-to-dust ratio $\Psi = 100$, a dust opacity at 3.3~mm $\kappa_\nu = 
		0.9$ cm$^2$ g$^{-1}$, and a dust temperature $T_{\rm dust} = 25$ K.
		\item[$^\mathrm{d}$] Includes JCMT fluxes from \citet{weintraub:1989}.
	\end{list}
   \end{table}	
   
   	\subsection{Opacity index}
	
	At low frequencies, i.e., in the Rayleigh-Jeans regime, the flux density, $F_\nu$, is
	related to frequency, $\nu$, by a power law: 
	$F_\nu \propto \nu^\alpha$. One can determine the dust opacity
	index, $\beta$, where $\kappa_\nu \propto \nu^\beta$, from the observed spectral index,
	$\alpha$, through
	\begin{equation}
		\beta \approx (\alpha - 2)(1 + \Delta),
	\end{equation}
	where $\Delta$ is the ratio of optically thick to optically thin emission from the
	disk \citep{beckwith:1990, beckwith:1991, rodmann:2006}.
	
	Due to the frequency dependence of the dust opacity, protoplanetary disks are generally
	optically thick at short wavelengths and optically thin at long wavelengths. Optically thick
	emission cannot be neglected in the inner disk where column densities get very high, even at 
	very long wavelengths. The ratio of optically thick
	to optically thin emission coming from the disk is given by
	\begin{equation}
		\Delta \approx - p \times \left[(2 - q) \ln \{(1 - p/2) \bar{\tau}\}\right]^{-1},
	\end{equation}
	where $p$ and $q$ are the disk's surface density and temperature indices, and $\bar{\tau}$
	is the average disk opacity at the frequency under consideration \citep[see][for details]{beckwith:1990}. Following 
	\citet{beckwith:1990} and \citet{rodmann:2006}, we used $p = 1.5$. $q$ is uniquely determined from
	the spectral index at wavelengths where the dust opacity is high ($\lambda \leq 100 \mu$m),
	and from IRAS photometry we obtained values ranging from 0.4 to 0.7. 
	$\bar{\tau}$ can in principle only be determined if the physical disk radius is known. 
	\citet{rodmann:2006} use $\bar{\tau} = 0.01$ at 7~mm. Taking $\beta = 1$ as a fiducial value for the opacity 
	index, we adopt $\bar{\tau} = 0.02$ at 3.3~mm and find values for $\Delta$ ranging from 0.18 to 0.22.
	
\begin{figure}
 \begin{flushleft}
  \begin{center}
    	\vspace{-0.1cm}
    	\hspace{0cm}
    	\hbox{\psfig{figure=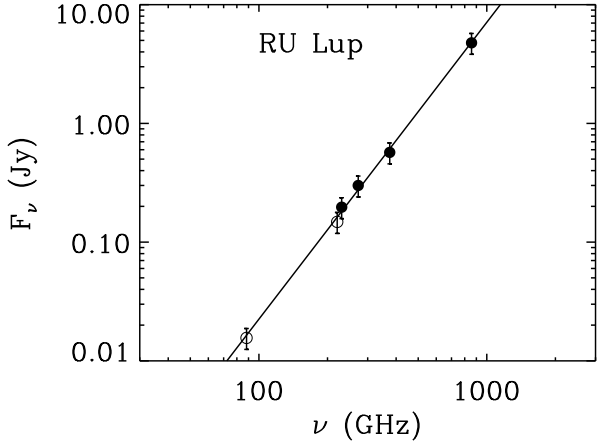,width=8.3cm}}
    	\hfill\parbox[b]{9cm}{\caption[]{$F_\nu$ vs. $\nu$ for RU~Lup. Filled symbols show fluxes from single-dish observations
	\citep[JCMT and SEST,][]{weintraub:1989,nurnberger:1997}, open symbols show the values for the interferometric observations from 
	this work. All points fall on a line with $\alpha = 2.5 \pm 0.1$, where $F_\nu \propto \nu^\alpha$. 
	This suggests that there is no extended emission over the size scale of the primary beams of the different telescopes, as 
	may be the result of, e.g., a remnant envelope or ambient cloud material. Hence, no extended emission is filtered out in the 
	interferometric observations. }\label{fig: rulup}}
  \end{center}
 \end{flushleft}
\end{figure}

	The spectral index $\alpha$ and the opacity index $\beta$ were determined from the ATCA 3.3-mm fluxes and SEST 1.3-mm
	fluxes from the literature \citep{nurnberger:1997, henning:1993}. 
	The uncertainties in $\alpha$ and $\beta$
	are typically 0.5, due to both the large uncertainties in the absolute fluxes of the data points
	and to the relatively short wavelength range over which $\alpha$ was determined. For the Lupus sources,
	$\alpha$ was also determined with the SMA 1.4-mm fluxes included, giving consistent
	results. The robustness of the value for $\alpha$ is illustrated by Fig.~\ref{fig: rulup},
	which shows the integrated fluxes from this work and single-dish (sub)millimetre fluxes from the
	literature for RU~Lup \citep{weintraub:1989, nurnberger:1997}, along with the fitted
	slope: $\alpha = 2.5 \pm 0.1$. We thus find that the formal uncertainties in $\alpha$ and $\beta$ for the other sources are probably 
	overestimates. Unfortunately, RU~Lup is the only source for which currently observations at such short wavelengths are available.
	The low error in the fit to its data point illustrates that more observations, 
	preferably over a larger wavelength regime, will help to better constrain values for
	$\alpha$, and hence $\beta$.
	
\begin{figure}
 \begin{flushleft}
  \begin{center}
    	\vspace{-0.1cm}
    	\hspace{0cm}
    	\hbox{\psfig{figure=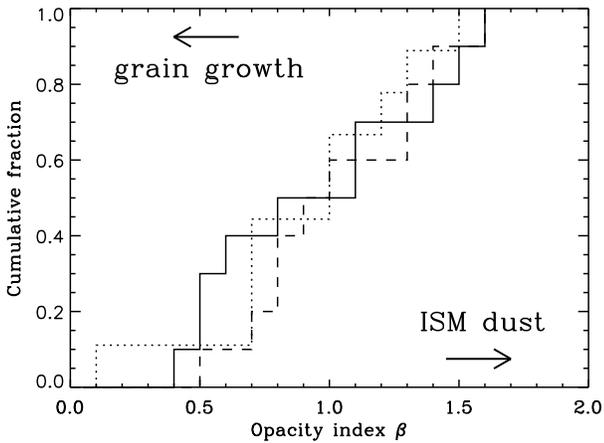,width=8.3cm}}
    	\hfill\parbox[b]{9cm}{\caption[]{
	Cumulative number of sources with an opacity index less than a given value of $\beta$ for the sources in our sample (solid line), those studied by 
	\citet{rodmann:2006} (dashed line), and those studied by \citet{natta:2004} (dotted line).}\label{fig: beta}}
  \end{center}
 \end{flushleft}
\end{figure}

	The values for $\alpha$ and $\beta$ for the sources in our sample are presented in Table~\ref{tab: Gaussian results}.
	Fig.~\ref{fig: beta} shows the cumulative fraction of sources with given dust-opacity indices
	for our sample and that of \citet{rodmann:2006}. \citeauthor{rodmann:2006} corrected for the contribution of free-free radiation at 7~mm. We estimate
	the contribution of free-free radiation to be $\sim$ 5\% at 3.3~mm, and hence it is not necessary to correct for
	it at this wavelength. Fig.~\ref{fig: rulup} illustrates that contamination from free-free emission is not an
	issue for RU~Lup. 
	The Kolmogorov-Smirnov test gives a
	probability of 58\% that \citeauthor{rodmann:2006}'s and our groups are drawn from the same distribution. 
	However, the uncertainties in $\beta$ are still quite large, and are not taken into account by the standard Kolmogorov-Smirnov test.
		
	\subsection{Disk masses}\label{sect: disk masses}
	
	For optically thin disks, the disk mass $M_{\rm disk}$ is directly proportianal to the flux $F_\nu$ \citep[see, e.g.,][]{hildebrand:1983, natta:2000}:
	\begin{equation}
		\label{eq: disk mass}
		M_{\rm disk} = \frac{F_\nu \Psi D^2}{\kappa_\nu B_\nu(T_{\rm dust})},
	\end{equation}
	where $\Psi$ is the gas-to-dust ratio, $D$ is the distance to the source, $\kappa_\nu$ is the dust opacity, and $B_\nu(T_{\rm dust})$ is the brightness at the frequency
	$\nu$ for a dust temperature $T_{\rm dust}$, as given by the Planck function. We used the integrated fluxes from the Gaussian fits,
	 and assumed $\Psi = 100$, $T_{\rm dust} = 25$ K, and $\kappa_\nu = 0.9$ cm$^2$ g$^{-1}$ 
	at 3.3~mm [cf. \citet[][]{beckwith:1990}\footnote{Note that \citet{beckwith:1990} estimate $\kappa_\nu = 0.1 (\nu/10^{12}$~Hz$)^\beta$ cm$^2$ g$^{-1}$, where 
	$\kappa_\nu$ is the
	opacity index {\it for the gas and the dust combined}, i.e., with an implied gas-to-dust ratio. Our values for $\kappa_\nu$, however, are for the dust alone, and
	hence we have to account for the gas-to-dust ratio explicitly.}]. 
	The derived disk masses for our sources range from $< 0.01$ up to 0.08~M$_\odot$, and are presented in Table~\ref{tab: Gaussian results}.
	Note that the mass estimates are quite uncertain and may easily be off by a factor of 2, due to the uncertainties in the parameter 
	values. Given the uncertainties
	in the measured quantities it is not meaningful to extend the analysis to include the contribution of optically thick emission and detailed 
	disk structure.
	
	\subsection{Molecular line emission}
	
\begin{figure}
 \begin{flushleft}
  \begin{center}
    	\vspace{-0.1cm}
    	\hspace{0cm}
    	\hbox{\psfig{figure=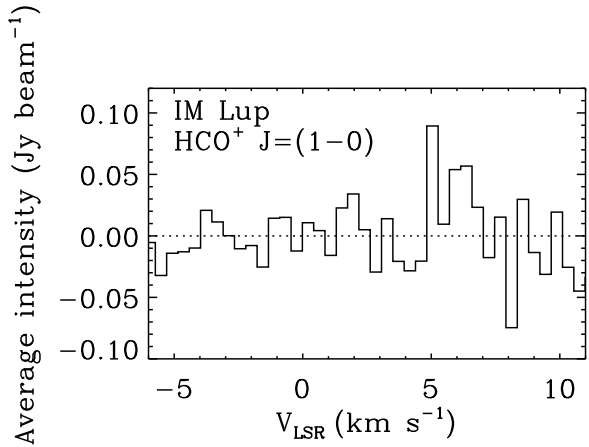,width=8.3cm}}
    	\hfill\parbox[b]{9cm}{\caption[]{Spectrum of HCO$^+$ $J$ = 1--0 line emission observed from IM Lup, at the position of the peak continuum emission, binned to 0.44 km 
	s$^{-1}$ velocity resolution, in the $2\farcs3 \times 1\farcs7$ synthesized beam.}\label{fig: line}}
  \end{center}
 \end{flushleft}
\end{figure}

	Both IM~Lup and WW~Cha were observed in spectral-line mode to search for HCO$^+$ $J$ = 1--0 emission. In general, CO is a more easily detectable gas tracer in the 
	millimetre regime, but 
	since the ATCA does not presently have the capability to observe at frequencies above 106~GHz, HCO$^+$ was used to investigate the gas component of disks. We 
	tentatively detect HCO$^+$ at $\sim 0.05$ Jy beam$^{-1}$ in IM~Lup, as presented in Fig.~\ref{fig: line}. 
	Van Kempen et al. (2006) \nocite{vankempen:2006} detected a double-peaked feature, consistent with a rotating gas disk, 
	using JCMT observations of the $^{12}$CO $J$ = 3--2 line. However, they
	find the line at $\sim$1 km~s$^{-1}$ lower velocity. The cause of this discrepancy is unclear.
	
	A ray-tracing programme \citep{hogerheijde:2000} was used to compute the line profile of HCO$^+$ $J$ = 1--0, using the model for IM~Lup 
	described in \citet{vankempen:2006}. The abundance of HCO$^+$ was set to be $10^{-8}$ with respect to H$_2$ in areas with temperatures above 30~K, and $10^{-12}$ in 
	areas with temperatures below 30~K. 
	An intensity of $\sim 0.02$ Jy beam$^{-1}$ was predicted, consistent with the observations within the uncertainties.
	
	No HCO$^+$ line emission is detected to a limit of 97 mJy beam$^{-1}$ (3$\sigma$) in the direction of WW~Cha. Combining our new results with those of \citet{wilner:2003} 
	who detected HCO$^+$ in \object{TW Hya}, but not in 
	\object{HD 100546}, we have two detections of HCO$^+$ and two non-detections. 
	Data for more targets are required, in order to determine whether the presence of HCO$^+$ is related to the disk's evolutionary state.
   
\section{Discussion and interpretations}\label{sect: discussion}

      \subsection{Grain growth}
      
      About half the sources that are detected are spatially resolved. The inferred physical sizes (disk radii of $\sim 100$~AU) indicate that the emission coming from the disks 
      is predominantly optically thin at millimetre wavelengths. This is illustrated by the work of
      \citet{testi:2001}, who do detailed modeling of the Herbig Ae stars \object{UX Ori} and \object{CQ Tau}, and find that disk radii of $\sim 100$~AU, combined with millimetre
      slopes of $\lesssim 3$ and 1.3~mm fluxes of $\sim 10^2$~mJy, are well explained by an opacity index $\beta = 1.0$. If we assume that all
      sources in our sample are optically thin at millimetre wavelengths, the opacity index $\beta$ can be determined. Seven out of ten detected sources have $\beta 
      \lesssim 1$, which can 
      be naturally explained by grain growth to millimetre and centimetre sizes \citep{draine:2006}. 
      
      \subsection{Comparison with Spitzer infrared data}
      
      Our values for $\beta$ indicate grain growth to sizes of millimetres and larger in the outer disks. The surface layers of the inner disks can be probed by infrared 
      observations and, as noted in Sect.~\ref{sect: introduction}, the 10-$\mu$m feature indicates the growth of grains through changes in its strength and shape.
      
\begin{figure}
 \begin{flushleft}
  \begin{center}
    	\vspace{-0.1cm}
    	\hspace{0cm}
    	\hbox{\hspace{-0.5cm}\psfig{figure=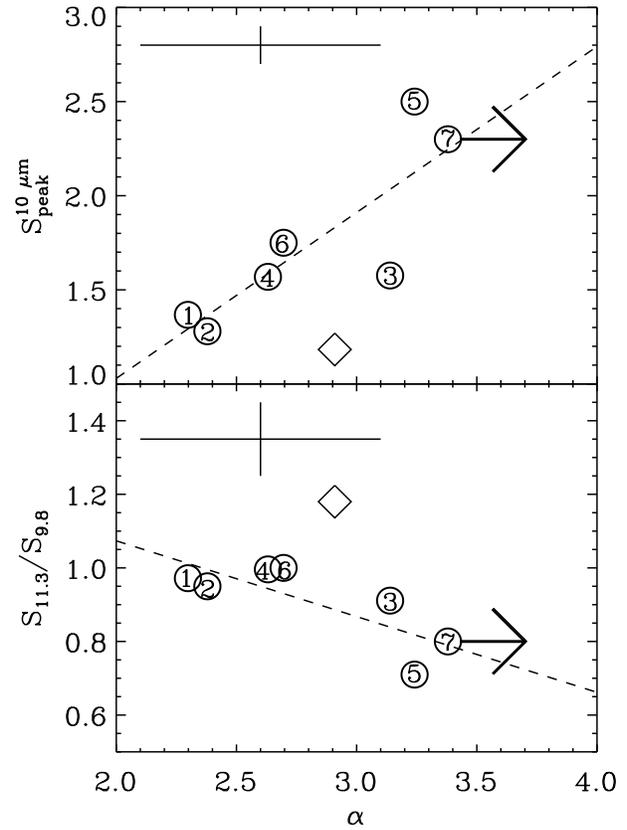,width=8.3cm}}
  	\hfill\parbox[b]{9cm}{\caption[]{Slope in the millimetre regime vs. the peak 10-$\mu$m flux (upper panel) and the ratio of the 11.3 to 9.8 $\mu$m flux (lower panel) 
	of the 10-$\mu$m silicate 
	feature, for those sources in our sample for which data are available. 1=HT~Lup, 2=GW~Lup, 3=IM~Lup, 4=RU~Lup, 5=CR~Cha, 6=WW~Cha, 7=Glass~I. The source T~Cha is 
	depicted with a diamond.}\label{fig: alpha vs. 10 mu}}
  \end{center}
 \end{flushleft}
\end{figure}

      In Fig.~\ref{fig: alpha vs. 10 mu} we compare the millimetre slope $\alpha$ as derived in Sect.~\ref{sect: results} with the ``strength'' or peak 10-$\mu$m flux 
      ($S^{\rm 10 \mu m}_{\rm peak}$) and to the ``shape'' or the ratio of the 11.3 to 9.8 $\mu$m flux ($S_{\rm 11.3}/S_{\rm 9.8}$) of the 10-$\mu$m silicate feature for those 
      sources in our sample that overlap with the samples of \citet{przygodda:2003} and \citet{kessler-silacci:2006}. Here the normalized 10-$\mu$m spectra $S_\nu$ are given by
      \begin{equation}
      	S_\nu = 1 + \frac{(F_\nu - F_{\nu, c})}{<F_{\nu, c}>},
      \end{equation}
      where $F_\nu$ is the observed spectrum and $F_{\nu, c}$ is the fitted continuum, and $<F_{\nu, c}>$ is the frequency-averaged continuum flux 
      (see Kessler-Silacci et al. 2006 for details).
      We find a positive correlation between $\alpha$ and $S^{\rm 10 \mu m}_{\rm peak}$ and a
      negative correlation between $\alpha$ and $S_{\rm 11.3}/S_{\rm 9.8}$, especially if we leave out the source T~Cha, which shows emission from polycyclic aromatic
      hydrocarbons at 11.3 $\mu$m and may be considerably older than the other sources in the sample. 
      As grains grow from submicron sizes to several microns, the 10-$\mu$m feature becomes weaker and less peaked \citep[see, e.g.,][]{kessler-silacci:2006},
      and thus
      the correlations we find may well indicate that once grain growth starts, the grains quickly grow from (sub)micron sizes to millimetre sizes and larger, 
      both in the inner and outer disks. This is consistent with grain-growth models \citep[][]{weidenschilling:1988,
      weidenschilling:1997, dullemond:2004b, dullemond:2005}, which show that growth to metre sizes is rapid ($\sim 10^3$~yr at 1~AU; $\sim 10^5$~yr at 30~AU).
      
      On the other hand, while a significant fraction of the dust has already grown to sizes of millimetres and centimetres, the infrared data indicate that micron-sized 
      grains are also still present, at least in the surface layers. This
      may be explained by assuming that not only grain growth, but also aggregate fragmentation takes place in the disks \citep[][]{dullemond:2005}. After about $10^4$ years a
      semi-stationary state is reached for sizes below $a \lesssim 1$~cm, which may last for several $10^6$ years. 
      
      Note that Fig.~\ref{fig: alpha vs. 10 mu} shows a trend rather than a bimodal distribution. \citet{dullemond:2005} suggest 
      fragmentation of grains to allow for the
      semi-stationary state that is observed in the disks around T Tauri stars.
      If the correlation between the 10-$\mu$m feature and the millimetre slope
      is confirmed by a more extensive dataset, it would indicate
      that when aggregates are fragmented in collisions, the size of the fragments increases as the aggregate sizes increase. Hence, when the 
      largest particles 
      grow from millimetre to centimetre sizes, the submicron-sized grains are no longer replenished, and the 10-$\mu$m feature -- tracing the 
      upper layers of the disk where
      the stirred-up small particles reside -- flattens while the millimetre slope -- which traces the midplane where the largest particles are 
      present -- becomes shallower.
      
      We also compared $\alpha$ to
      the spectral index over the 13 to 35 $\mu$m range from \citet{kessler-silacci:2006}. However, $\alpha$ could only be determined for five of the sources in 
      \citeauthor{kessler-silacci:2006}, and no obvious correlation is found for this small sample.

      \subsection{Comparison with Herbig Ae/Be stars}
      
      It is interesting to note that \citet{acke:2004b} do {\it not} find a correlation between the 10-$\mu$m silicate feature and the (sub)millimetre spectral index for 
      their sample of 26 Herbig Ae/Be stars. On the other hand, \citet{acke:2004a} {\it do} find a correlation between the shape of the mid-IR (12--60~$\mu$m) SED and the 
      (sub)millimetre slope in their sample. They interpret this finding as a correlation between disk geometry and grain size: as grains grow, the disk structure evolves from 
      flared to geometrically flat \citep[see, e.g.,][]{dullemond:2002}. For the twelve sources in our sample for which it was possible to
      determine 
      $\alpha$, we determined the non-colour-corrected {\it IRAS}
      [12]--[60] colour and compared this to $\alpha$. No clear correlation was found, consistent with the above lack of correlation with the 13--35~$\mu$m spectral index. 
      This may indicate that grain growth has a less dramatic effect on the geometry
      of disks around the low-mass T Tauri stars than on the intermediate-mass Herbig Ae/Be stars, or that {\it IRAS} fluxes are contaminated by other (extended) emission 
      in the large {\it IRAS} beams. 
      
      \citet{natta:2004} did a study similar to ours for a sample of six isolated Herbig Ae stars and three T Tauri stars. They estimated $\beta$ in two different ways, 
      firstly using only interferometric data, and secondly with single-dish data included. Their results are included in Fig.~\ref{fig: beta}, where the values 
      for $\beta$ with the
      single-dish data included were used for comparison. 
      According to the Kolmogorov-Smirnov test there is a 65\% probability that the sample of \citeauthor{natta:2004} and our sample are drawn from the same distribution.
      It is interesting to note that in general the sources in \citeauthor{natta:2004}'s sample are considerably older than those sources in the current work and that of 
      \citet{rodmann:2006}.
      This may be an extra indication that the size distribution of grains in protoplanetary disks remains stationary for several Myr once sizes of millimetres are reached.
      
\section{Concluding remarks}\label{sect: conclusions}

      We used the ATCA to make interferometric observations of fifteen southern T Tauri sources at 3.3~mm. Ten of the sources are located in Chamaeleon, the remaining five in 
      Lupus. The Lupus sources were also observed at 1.4~mm with the SMA. The main results are as follows.
      \begin{itemize}
      	\item Five of the Chamaeleon sources were detected at the 3$\sigma$ level or better with the ATCA, for the other five sources we have strict upper limits.
	\item All five Lupus sources were detected at better than 3$\sigma$ with the ATCA, and at better than 15$\sigma$ with the SMA.
	\item Five of the ten detected sources are spatially resolved with the ATCA, and two of the five Lupus sources are resolved with the SMA.
	This indicates that the emission coming from these sources is optically thin.
	\item Adopting optically thin emission for all sources in our sample, we estimated the opacity index $\beta$. We find that $\beta \lesssim 1$ for seven of
	the ten detected sources. Such values for $\beta$ can be naturally explained by grain growth up to millimetre sizes and beyond.
	\item We find a tentative correlation between the millimetre slope and the peak 10-$\mu$m flux and ratio of the 11.3 to 9.8 $\mu$m flux of the 10-$\mu$m silicate feature. 
	This indicates that grain growth takes place in the outer disk and in the surface layers of the inner disk simultaneously. It also confirms earlier findings that 
	agglomerate destruction must take place in the disk to preserve the
	small grain population. If confirmed by more extensive datasets,
	this correlation may be an indication that larger aggregates produce
	larger particles when fragmented in collisions.
	\item HCO$^+$ was tentatively detected in IM~Lup, and was not seen in WW~Cha.
      \end{itemize}
      This type of work will be greatly assisted by the upcoming 7-mm upgrade for the ATCA, as well as by the installation of a new correlator via the Compact Array
      broadband backend project. Estimating the millimetre slope over a wavelength range up to 7~mm will reduce the error in $\alpha$ -- and hence in $\beta$ -- from 
      $\sim 0.5$ to $\sim 0.2$. At these longer wavelengths, however, the contribution of free-free emission becomes important, and lower-frequency measurements are
      required to separate the various contributions. A pilot study to determine the contribution of free-free emission at centimetre wavelengths in our sources was carried out 

      The capability to observe the continuum across a 2~GHz bandwidth, as opposed to the current 
      128~MHz, will provide a marked increase in sensitivity, allowing more disks to be detected. Furthermore, as the new generation of (sub)millimetre interferometers, notably 
      eSMA, CARMA, and ALMA, becomes available to the scientific community, it becomes possible to resolve the circumstellar disks at millimetre and submillimetre wavelengths 
      down to subarcsecond scales. A survey of a large 
      sample of objects, in different clouds, and at higher resolution will greatly enhance our understanding of the timescale for grain growth and the building of planetary 
      systems as a function of disk radius.
      
\begin{acknowledgements}
      We would like to thank the ATNF for their hospitality and assistance, and specifically Tony Wong for extensive assistance during the 
      observations and the data reduction. SMA staff, in particular Alison Peck, are thanked for scheduling observations of the Lupus sources as a
      filler programme and for carrying out the observations. Partial support for this work was provided by a Netherlands Research School For Astronomy network 2 grant, and by an 
      Netherlands Organisation for Scientific Research Spinoza grant. CMW acknowledges financial support from an ARC Australian Research Fellowship. We are grateful to Jackie 
      Kessler-Silacci for providing us with infrared data, and to Tim van Kempen for useful discussions on IM~Lup and calculating the HCO$^+$ emission in his model.
      Finally, we would like to thank our referee, Claire Chandler, for useful comments that significantly improved this paper.
\end{acknowledgements}

\bibliographystyle{aa}
\bibliography{references}

\begin{thebibliography}{56}
\expandafter\ifx\csname natexlab\endcsname\relax\def\natexlab#1{#1}\fi

\bibitem[{{Acke} \& {van den Ancker}(2004)}]{acke:2004b}
{Acke}, B. \& {van den Ancker}, M.~E. 2004, \aap, 426, 151

\bibitem[{{Acke} {et~al.}(2004){Acke}, {van den Ancker}, {Dullemond}, {van
  Boekel}, \& {Waters}}]{acke:2004a}
{Acke}, B., {van den Ancker}, M.~E., {Dullemond}, C.~P., {van Boekel}, R., \&
  {Waters}, L.~B.~F.~M. 2004, \aap, 422, 621

\bibitem[{{Alcala} {et~al.}(1995){Alcala}, {Krautter}, {Schmitt}, {Covino},
  {Wichmann}, \& {Mundt}}]{alcala:1995}
{Alcala}, J.~M., {Krautter}, J., {Schmitt}, J.~H.~M.~M., {et~al.} 1995, \aaps,
  114, 109

\bibitem[{{Beckwith} \& {Sargent}(1991)}]{beckwith:1991}
{Beckwith}, S.~V.~W. \& {Sargent}, A.~I. 1991, \apj, 381, 250

\bibitem[{{Beckwith} {et~al.}(1990){Beckwith}, {Sargent}, {Chini}, \&
  {Guesten}}]{beckwith:1990}
{Beckwith}, S.~V.~W., {Sargent}, A.~I., {Chini}, R.~S., \& {Guesten}, R. 1990,
  \aj, 99, 924

\bibitem[{{Bouwman} {et~al.}(2001){Bouwman}, {Meeus}, {de Koter}, {Hony},
  {Dominik}, \& {Waters}}]{bouwman:2001}
{Bouwman}, J., {Meeus}, G., {de Koter}, A., {et~al.} 2001, \aap, 375, 950

\bibitem[{{Calvet} {et~al.}(2002){Calvet}, {D'Alessio}, {Hartmann}, {Wilner},
  {Walsh}, \& {Sitko}}]{calvet:2002}
{Calvet}, N., {D'Alessio}, P., {Hartmann}, L., {et~al.} 2002, \apj, 568, 1008

\bibitem[{{Chen} {et~al.}(1997){Chen}, {Grenfell}, {Myers}, \&
  {Hughes}}]{chen:1997}
{Chen}, H., {Grenfell}, T.~G., {Myers}, P.~C., \& {Hughes}, J.~D. 1997, \apj,
  478, 295

\bibitem[{{Chiang} \& {Goldreich}(1999)}]{chiang:1999}
{Chiang}, E.~I. \& {Goldreich}, P. 1999, \apj, 519, 279

\bibitem[{{Comer\'{o}n}(2006)}]{comeron:2006}
{Comer\'{o}n}, F. 2006, in 'Handbook of star forming regions', ed. B. Reipurth,
  ASP Conf. Ser., in press (ESO press)

\bibitem[{{Comer{\'o}n} {et~al.}(1999){Comer{\'o}n}, {Rieke}, \&
  {Neuh{\"a}user}}]{comeron:1999}
{Comer{\'o}n}, F., {Rieke}, G.~H., \& {Neuh{\"a}user}, R. 1999, \aap, 343, 477

\bibitem[{{D'Alessio} {et~al.}(2006){D'Alessio}, {Calvet}, {Hartmann},
  {Franco-Hern{\'a}ndez}, \& {Serv{\'{\i}}n}}]{dalessio:2006}
{D'Alessio}, P., {Calvet}, N., {Hartmann}, L., {Franco-Hern{\'a}ndez}, R., \&
  {Serv{\'{\i}}n}, H. 2006, \apj, 638, 314

\bibitem[{{Dominik} {et~al.}(2006){Dominik}, {Blum}, {Cuzzi}, \&
  {Wurm}}]{dominik:2006}
{Dominik}, C., {Blum}, J., {Cuzzi}, J.~N., \& {Wurm}, G. 2006, in Proceedings
  of PPV, B. Reipurth, D. Jewitt, and K. Keil eds., astro-ph/0602617 (the
  University of Arizona Press)

\bibitem[{{Draine}(2006)}]{draine:2006}
{Draine}, B.~T. 2006, \apj, 636, 1114

\bibitem[{{Dullemond}(2002)}]{dullemond:2002}
{Dullemond}, C.~P. 2002, \aap, 395, 853

\bibitem[{{Dullemond} \& {Dominik}(2004{\natexlab{a}})}]{dullemond:2004}
{Dullemond}, C.~P. \& {Dominik}, C. 2004{\natexlab{a}}, \aap, 417, 159

\bibitem[{{Dullemond} \& {Dominik}(2004{\natexlab{b}})}]{dullemond:2004b}
{Dullemond}, C.~P. \& {Dominik}, C. 2004{\natexlab{b}}, \aap, 421, 1075

\bibitem[{{Dullemond} \& {Dominik}(2005)}]{dullemond:2005}
{Dullemond}, C.~P. \& {Dominik}, C. 2005, \aap, 434, 971

\bibitem[{{Evans} {et~al.}(2003){Evans}, {Allen}, {Blake}, {Boogert}, {Bourke},
  {Harvey}, {Kessler}, {Koerner}, {Lee}, {Mundy}, {Myers}, {Padgett},
  {Pontoppidan}, {Sargent}, {Stapelfeldt}, {van Dishoeck}, {Young}, \&
  {Young}}]{evans:2003}
{Evans}, N.~J., {Allen}, L.~E., {Blake}, G.~A., {et~al.} 2003, \pasp, 115, 965

\bibitem[{{Gauvin} \& {Strom}(1992)}]{gauvin:1992}
{Gauvin}, L.~S. \& {Strom}, K.~M. 1992, \apj, 385, 217

\bibitem[{{G{\"u}rtler} {et~al.}(1999){G{\"u}rtler}, {Schreyer}, {Henning},
  {Lemke}, \& {Pfau}}]{gurtler:1999}
{G{\"u}rtler}, J., {Schreyer}, K., {Henning}, T., {Lemke}, D., \& {Pfau}, W.
  1999, \aap, 346, 205

\bibitem[{{Henize} \& {Mendoza}(1973)}]{henize:1973}
{Henize}, K.~G. \& {Mendoza}, E.~E. 1973, \apj, 180, 115

\bibitem[{{Henning} {et~al.}(1993){Henning}, {Pfau}, {Zinnecker}, \&
  {Prusti}}]{henning:1993}
{Henning}, T., {Pfau}, W., {Zinnecker}, H., \& {Prusti}, T. 1993, \aap, 276,
  129

\bibitem[{{Herbig} \& {Bell}(1988)}]{herbig:1988}
{Herbig}, G.~H. \& {Bell}, K.~R. 1988, {Catalog of emission line stars of the
  orion population : 3 : 1988} (Lick Observatory Bulletin, Santa Cruz: Lick
  Observatory, |c1988)

\bibitem[{{Herbig} \& {Kameswara Rao}(1972)}]{herbig:1972}
{Herbig}, G.~H. \& {Kameswara Rao}, N. 1972, \apj, 174, 401

\bibitem[{{Hildebrand}(1983)}]{hildebrand:1983}
{Hildebrand}, R.~H. 1983, \qjras, 24, 267

\bibitem[{{Ho} {et~al.}(2004){Ho}, {Moran}, \& {Lo}}]{ho:2004}
{Ho}, P.~T.~P., {Moran}, J.~M., \& {Lo}, K.~Y. 2004, \apjl, 616, L1

\bibitem[{{Hogerheijde} \& {van der Tak}(2000)}]{hogerheijde:2000}
{Hogerheijde}, M.~R. \& {van der Tak}, F.~F.~S. 2000, \aap, 362, 697

\bibitem[{{Hughes} {et~al.}(1994){Hughes}, {Hartigan}, {Krautter}, \&
  {Kelemen}}]{hughes:1994}
{Hughes}, J., {Hartigan}, P., {Krautter}, J., \& {Kelemen}, J. 1994, \aj, 108,
  1071

\bibitem[{{J{\o}rgensen} {et~al.}(2005){J{\o}rgensen}, {Bourke}, {Myers},
  {Sch{\"o}ier}, {van Dishoeck}, \& {Wilner}}]{jorgensen:2005}
{J{\o}rgensen}, J.~K., {Bourke}, T.~L., {Myers}, P.~C., {et~al.} 2005, \apj,
  632, 973

\bibitem[{{Kenyon} \& {G{\'o}mez}(2001)}]{kenyon:2001}
{Kenyon}, S.~J. \& {G{\'o}mez}, M. 2001, \aj, 121, 2673

\bibitem[{{Kessler-Silacci} {et~al.}(2006){Kessler-Silacci}, {Augereau},
  {Dullemond}, {Geers}, {Lahuis}, {Evans}, {van Dishoeck}, {Blake}, {Boogert},
  {Brown}, {J{\o}rgensen}, {Knez}, \& {Pontoppidan}}]{kessler-silacci:2006}
{Kessler-Silacci}, J., {Augereau}, J.-C., {Dullemond}, C.~P., {et~al.} 2006,
  \apj, 639, 275

\bibitem[{{Lawson} {et~al.}(1996){Lawson}, {Feigelson}, \&
  {Huenemoerder}}]{lawson:1996}
{Lawson}, W.~A., {Feigelson}, E.~D., \& {Huenemoerder}, D.~P. 1996, \mnras,
  280, 1071

\bibitem[{{Malfait} {et~al.}(1998){Malfait}, {Waelkens}, {Waters},
  {Vandenbussche}, {Huygen}, \& {de Graauw}}]{malfait:1998}
{Malfait}, K., {Waelkens}, C., {Waters}, L.~B.~F.~M., {et~al.} 1998, \aap, 332,
  L25

\bibitem[{{Meeus} {et~al.}(2001){Meeus}, {Waters}, {Bouwman}, {van den Ancker},
  {Waelkens}, \& {Malfait}}]{meeus:2001}
{Meeus}, G., {Waters}, L.~B.~F.~M., {Bouwman}, J., {et~al.} 2001, \aap, 365,
  476

\bibitem[{{Natta} {et~al.}(2000){Natta}, {Grinin}, \& {Mannings}}]{natta:2000}
{Natta}, A., {Grinin}, V., \& {Mannings}, V. 2000, Protostars and Planets IV,
  559

\bibitem[{{Natta} {et~al.}(2004){Natta}, {Testi}, {Neri}, {Shepherd}, \&
  {Wilner}}]{natta:2004}
{Natta}, A., {Testi}, L., {Neri}, R., {Shepherd}, D.~S., \& {Wilner}, D.~J.
  2004, \aap, 416, 179

\bibitem[{{Nomura} \& {Nakagawa}(2006)}]{nomura:2006}
{Nomura}, H. \& {Nakagawa}, Y. 2006, \apj, 640, 1099

\bibitem[{{N{\"u}rnberger} {et~al.}(1997){N{\"u}rnberger}, {Chini}, \&
  {Zinnecker}}]{nurnberger:1997}
{N{\"u}rnberger}, D., {Chini}, R., \& {Zinnecker}, H. 1997, \aap, 324, 1036

\bibitem[{{Pontoppidan} {et~al.}(2003){Pontoppidan}, {Fraser}, {Dartois},
  {Thi}, {van Dishoeck}, {Boogert}, {d'Hendecourt}, {Tielens}, \&
  {Bisschop}}]{pontoppidan:2003}
{Pontoppidan}, K.~M., {Fraser}, H.~J., {Dartois}, E., {et~al.} 2003, \aap, 408,
  981

\bibitem[{{Przygodda} {et~al.}(2003){Przygodda}, {van Boekel},
  {{\`A}brah{\`a}m}, {Melnikov}, {Waters}, \& {Leinert}}]{przygodda:2003}
{Przygodda}, F., {van Boekel}, R., {{\`A}brah{\`a}m}, P., {et~al.} 2003, \aap,
  412, L43

\bibitem[{{Rodgers} {et~al.}(1978){Rodgers}, {Peterson}, \&
  {Harding}}]{rodgers:1978}
{Rodgers}, A.~W., {Peterson}, B.~A., \& {Harding}, P. 1978, \apj, 225, 768

\bibitem[{{Rodmann} {et~al.}(2006){Rodmann}, {Henning}, {Chandler}, {Mundy}, \&
  {Wilner}}]{rodmann:2006}
{Rodmann}, J., {Henning}, T., {Chandler}, C.~J., {Mundy}, L.~G., \& {Wilner},
  D.~J. 2006, \aap, 446, 211

\bibitem[{{Safronov} \& {Zvjagina}(1969)}]{safronov:1969}
{Safronov}, V.~S. \& {Zvjagina}, E.~V. 1969, Icarus, 10, 109

\bibitem[{{Sault} {et~al.}(1995){Sault}, {Teuben}, \& {Wright}}]{sault:1995}
{Sault}, R.~J., {Teuben}, P.~J., \& {Wright}, M.~C.~H. 1995, in ASP Conf. Ser.
  77: Astronomical Data Analysis Software and Systems IV, 433

\bibitem[{{Testi} {et~al.}(2001){Testi}, {Natta}, {Shepherd}, \&
  {Wilner}}]{testi:2001}
{Testi}, L., {Natta}, A., {Shepherd}, D.~S., \& {Wilner}, D.~J. 2001, \apj,
  554, 1087

\bibitem[{{van Boekel} {et~al.}(2005){van Boekel}, {Min}, {Waters}, {de Koter},
  {Dominik}, {van den Ancker}, \& {Bouwman}}]{vanboekel:2005}
{van Boekel}, R., {Min}, M., {Waters}, L.~B.~F.~M., {et~al.} 2005, \aap, 437,
  189

\bibitem[{{van Boekel} {et~al.}(2003){van Boekel}, {Waters}, {Dominik},
  {Bouwman}, {de Koter}, {Dullemond}, \& {Paresce}}]{vanboekel:2003}
{van Boekel}, R., {Waters}, L.~B.~F.~M., {Dominik}, C., {et~al.} 2003, \aap,
  400, L21

\bibitem[{{van den Ancker} {et~al.}(1998){van den Ancker}, {de Winter}, \&
  {Tjin A Djie}}]{vandenancker:1998}
{van den Ancker}, M.~E., {de Winter}, D., \& {Tjin A Djie}, H.~R.~E. 1998,
  \aap, 330, 145

\bibitem[{{van Kempen} {et~al.}(2006){van Kempen}, {van Dishoeck}, {Brinch}, \&
  {Hogerheijde}}]{vankempen:2006}
{van Kempen}, T.~A., {van Dishoeck}, E.~F., {Brinch}, C., \& {Hogerheijde},
  M.~R. 2006, A\&A accepted, astro-ph/0606302

\bibitem[{{Weidenschilling}(1988)}]{weidenschilling:1988}
{Weidenschilling}, S.~J. 1988, {Formation processes and time scales for
  meteorite parent bodies} (Meteorites and the Early Solar System), 348--371

\bibitem[{{Weidenschilling}(1997)}]{weidenschilling:1997}
{Weidenschilling}, S.~J. 1997, Icarus, 127, 290

\bibitem[{{Weintraub} {et~al.}(1989){Weintraub}, {Sandell}, \&
  {Duncan}}]{weintraub:1989}
{Weintraub}, D.~A., {Sandell}, G., \& {Duncan}, W.~D. 1989, \apjl, 340, L69

\bibitem[{{Whittet} {et~al.}(1997){Whittet}, {Prusti}, {Franco}, {Gerakines},
  {Kilkenny}, {Larson}, \& {Wesselius}}]{whittet:1997}
{Whittet}, D.~C.~B., {Prusti}, T., {Franco}, G.~A.~P., {et~al.} 1997, \aap,
  327, 1194

\bibitem[{{Wilner} {et~al.}(2003){Wilner}, {Bourke}, {Wright}, {J{\o}rgensen},
  {van Dishoeck}, \& {Wong}}]{wilner:2003}
{Wilner}, D.~J., {Bourke}, T.~L., {Wright}, C.~M., {et~al.} 2003, \apj, 596,
  597

\bibitem[{{Wilner} {et~al.}(2000){Wilner}, {Ho}, {Kastner}, \&
  {Rodr{\'{\i}}guez}}]{wilner:2000}
{Wilner}, D.~J., {Ho}, P.~T.~P., {Kastner}, J.~H., \& {Rodr{\'{\i}}guez}, L.~F.
  2000, \apjl, 534, L101

\end{thebibliography}

\end{document}